\documentclass[prd,showpacs,showkeys,superscriptaddress,nofootinbib,twocolumn,floatfix]{revtex4}
\usepackage{amsfonts}
\usepackage{graphicx,color,amsmath,amssymb}
\usepackage{subfigure}

\def\blfootnote{\xdef\@thefnmark{}\@footnotetext}
\begin{document}

\title{Second-Order Dissociation and Transition of Heavy Quarkonia in the Quark-Gluon Plasma}
\author{Shouxing Zhao}
\affiliation{Department of Applied Physics, Nanjing University of Science and Technology, Nanjing 210094, China}
\author{Min He}
\affiliation{Department of Applied Physics, Nanjing University of Science and Technology, Nanjing 210094, China}
\affiliation{Shanghai Research Center for Theoretical Nuclear Physics, NSFC and Fudan University, Shanghai 200438, China}

\date{\today}

\begin{abstract}
We revisit the dissociation of heavy quarkonia by thermal partons at the next-to-leading order (NLO, also known as inelastic parton scattering dissociation) in the Quark-Gluon Plasma (QGP). Utilizing the chromo-electric dipole coupling from QCD multipole expansion as an effective Hamiltonian, this has been conducted in the approach of second-order quantum mechanical perturbation theory, which allows us to systematically incorporate the bound state wave functions. Employing the quarkonium wave functions and binding energies obtained from an in-medium potential model, we then numerically evaluate the dissociation cross sections and rates for various charmonia and bottomonia, where the infrared and collinear divergences are regularized by the thermal masses of medium partons. We demonstrate that distinct from the leading order (LO, also known as gluo-dissociation) counterparts peaking at relatively low gluon energy and falling off thereafter, the NLO cross sections first grow and then nearly saturate as the incident parton energy increases, as a result of the outgoing parton carrying away the excess energy. The resulting NLO dissociation rates increase with temperature and take over from the LO counterparts toward high temperatures, similar to pertinent findings from previous studies. We also evaluate the in-medium second-order transition between different bound states, which may contribute to the total thermal decay widths of heavy quarkonia in the QGP.

\end{abstract}


\maketitle

%%%%%%%%%%%%%%%%%%%%%%%%%%%%%%%%%%%%%%%%%%%%%%%%%%%%%%%%
\section{Introduction}
\label{sec_intro}
%%%%%%%%%%%%%%%%%%%%%%%%%%%%%%%%%%%%%%%%%%%%%%%%%%%%%%%

Heavy quarkonia, the bound states of charm or bottom quark-antiquark pair, represent an versatile laboratory for testing properties of strong interactions not only in vacuum but also at finite temperatures and densities such as the environment of the deconfined medium (known as Quark-Gluon Plasma, QGP) created in relativistic heavy-ion collisions~\cite{Brambilla:2010cs,Rapp:2008tf,Braun-Munzinger:2009dzl,Kluberg:2009wc,Mocsy:2013syh,Andronic:2015wma,Rothkopf:2019ipj,Zhao:2020jqu,Andronic:2024oxz}.
It was first advocated by Matsui and Satz~\cite{Matsui:1986dk} that the screening of color charges in the deconfined medium should weaken the binding between the heavy quark and its antiquark, leading eventually to the melting of their bound states and thus serving as a signal of the formation of QGP. This picture further suggests that different heavy quarkonium should melt sequentially at temperatures in the order of their vacuum binding energies~\cite{Digal:2001ue}.

However, the complexity of in-medium quarkonium dynamics prevents one from using them as a straightforward thermometer of the medium based on the above purely static screening scenario. On the one hand, mechanisms involving collisions with medium constituents can lead to dynamical dissociation of the bound states~\cite{Kharzeev:1994pz,Grandchamp:2001pf}. On the other hand, in situations where heavy quarks are abundantly produced, {\it e.g.}, charm production in Pb-Pb collisions at the LHC energy, individual heavy quarks and antiquarks diffusing and thermalizing in the QGP may recombine and regenerate the bound states~\cite{Braun-Munzinger:2000csl,Thews:2000rj,Grandchamp:2003uw}. The dissociation rates/collisional widths generated from partonic inelastic scatterings with thermal partons serve as a pivotal input for both semiclassical~\cite{ Zhao:2010nk,Song:2011nu,Strickland:2011mw,Zhou:2014kka} and quantum~\cite{Brambilla:2016wgg,Brambilla:2020qwo,Yao:2021lus} transport descriptions of the heavy quarkonium production in the QGP. These reaction rates, as facilitated by the principle of detailed balance and implemented in the kinetic rate equations~\cite{Zhao:2010nk,Wu:2024gil}, also govern the regeneration processes.

From the theoretical point of view, the dissociation of heavy quarkonium through collisions with thermal partons can be classified into leading order (LO) and next-to-leading order (NLO) processes. The LO process refers to the gluo-dissociation, $g+\Psi\rightarrow Q+\bar{Q}$ ($\Psi$ denotes a heavy quarkonium, and $Q$ ($\bar{Q}$) the heavy quark (antiquark)), in which the bound state absorbs a thermal gluon from the medium and breaks up into an unbound color octet, $(Q\bar{Q})_8$, by overcoming the binding energy. The analysis of this process was initiated by Peskin within the operator-product-expansion approach, who identified the coupling of the tightly bound heavy quarkonium with external soft gluons as a gauge invariant color-electric dipole type~\cite{Peskin:1979va,Bhanot:1979vb}. Using the same coupling or its variants ({\it e.g.}, nonrelativistic approximation of the associated Bethe-Salpeter amplitude~\cite{Oh:2001rm}, the color-magnetic dipole transition~\cite{Chen:2017jje}), many authors have thereafter investigated the same process~\cite{Kharzeev:1994pz,Oh:2001rm,Wong:2004zr,Brezinski:2011ju,Liu:2013kkg,Chen:2017jje}. The cross section of this process typically shows a peak at the incident gluon energy when the gluon wavelength matches the bound state size, but quickly drops off toward higher gluon energies~\cite{Chen:2017jje}.

Going to the NLO, $p+\Psi\rightarrow p+ Q+\bar{Q}$ (the thermal parton $p=g, q$ or $\bar q$), also known as dissociation by inelastic parton scattering, the outgoing parton in the final state carries away the excess energy, which thus may help maintain the efficiency for the break-up of the bound state even for incident gluons with higher energies. The phenomenological relevance of this process was first put forward from the ``quasifree" perspective~\cite{Grandchamp:2001pf}, by noting that in the limit of small binding (or large bound state size) at high temperatures, the incident thermal parton with energy $\sim T$ (temperature of the medium) ``sees" the individual constituents within the bound state, rather than the quarkonium as a whole, and therefore the scattering effectively occurs on a single $Q$ with the other $\bar{Q}$ being a spectator (or vice versa)~\cite{Riek:2010py}; the interference effect between the parton scattering off the $Q$ and $\bar{Q}$ as a result of the residual binding was originally ignored and later corrected for~\cite{Du:2017qkv} in the ``quasifree" treatment. The cross section of the NLO process was also computed in the perturbative QCD approach~\cite{Song:2005yd} (and recently updated in~\cite{Hong:2018vgp}), by identifying an effective vertex from the non-relativistic Bethe-Salpeter amplitude used earlier in the LO calculation~\cite{Oh:2001rm}.

The identification of an imaginary part of the heavy quark potential in the effective field theory (EFT) approach~\cite{Laine:2006ns,Beraudo:2007ky,Brambilla:2008cx,Brambilla:2013dpa}, as arising from the Landau damping of the space-like gluon exchanged between $Q$ and $\bar{Q}$ in the environment of QGP, represents a paradigm shift for the NLO process of heavy quarkonium dissociation. Microscopically, the Landau damping underlying the imaginary potential originates from the energy transfer from the low-frequency gluons mediating the interaction between $Q$ and $\bar{Q}$ to the ``hard" particles with energy of order $\sim T$ in the thermal bath~\cite{Laine:2006ns,Beraudo:2007ky}. This can be clearly seen from ``cutting" the (one-loop) self-energy diagram of the exchanged space-like gluon, thereby making a close connection to aforementioned ``quasifree" process as well as the elastic collisional broadening/thermalization of a single heavy quark. The imaginary static potential has been now verified from first-principle lattice QCD analysis of the Wilson loop spectral functions~\cite{Burnier:2014ssa,Bala:2021fkm,Bazavov:2023dci}.

It has been established that the LO process dominates the heavy quarkonium dissociation at low temperatures when the associated binding energy is still large compared to the screening mass, whereas the NLO takes over as the binding becomes weak toward high temperatures~\cite{Zhao:2010nk,Wu:2024gil,Du:2017qkv,Hong:2018vgp,Brambilla:2013dpa}. However, the dissociation rates currently used in different heavy quarkonium transport models suffer from large discrepancies~\cite{Andronic:2024oxz}, calling further for reliable evaluations of the heavy quarkonium dissociation processes. The purpose of the present work is to revisit the cross sections and rates for the NLO heavy quarkonium dissociation by thermal partons in the QGP from the dynamical scattering point of view, in the similar spirit of~\cite{Hong:2018vgp} but from different approach. We calculate the NLO inelastic scattering cross sections within the framework of the second-order quantum mechanical perturbation theory, utilizing the color-electric dipole coupling of the heavy quarkonium with external gluons~\cite{Peskin:1979va,Yan:1980uh,Brambilla:2004jw,Voloshin:2007dx,Sumino:2014qpa} as the effective interaction Hamiltonian. This approach was first introduced in~\cite{Chen:2018dqg} by one of the present authors as an attempt to study the NLO break-up of heavy quarkonium by thermal gluons. The present work aims to present a more systematic and clear treatment of the approach and generalize the calculation to include thermal quarks (and antiquarks). Another theme tackled in the present work is the transitions between different eigenstates of quarkonium bound states, which also occur at the second order involving two gluons, since the bound states in both of the initial and final state are color singlets. We compute the pertinent transition rates, which may contribute to the total thermal spectral width of heavy quarkonium as evaluated in lattice QCD~\cite{Larsen:2019zqv}. The current approach allows us to systematically incorporate the bound state wave functions, thus going beyond the ``quasifree" approximation. The direct calculation of the cross sections from dynamical scattering processes also makes the current approach distinct from the method in~\cite{Brambilla:2013dpa} where the NLO cross section was extracted from an formula of the thermal decay width expressed in terms of the imaginary potential~\cite{Laine:2006ns}. The color-electric dipole coupling was also employed in~\cite{Yao:2018sgn} to compute the heavy quarkonium scattering amplitudes in the QGP from the potential non-relativistic QCD (pNRQCD) perspective, but no derivations of the cross sections were given.

The article is organized as follows. In Sec.~\ref{sec_hamiltonian}, the QCD system of a heavy quarkonium and external light quarks and gluons is formulated in terms of an effective Hamiltonian. In Sec.~\ref{sec_dissociation}, we derive the cross sections for the NLO dissociation of heavy quarkonium by thermal gluons and quarks within the second-order quantum mechanical perturbation framework. In Sec.~\ref{sec_numerical_diss}, we employ the bound state wave functions and binding energies computed from an in-medium potential model and numerically evaluate the NLO cross sections and dissociation rates for various kinds of charmonia and bottomonia. In Sec.~\ref{sec_transition}, we study the second-order transition processes between different eigenstates of quarkonium singlets and compare the pertinent cross sections and rates to the NLO dissociation counterparts. We finally summarize in Sec.~\ref{sec_sum}.

%%%%%%%%%%%%%%%%%%%%%%%%%%%%%%%%%%%%%%%%%%%%%%%%%%%%%%%%%%%%%%%%%%%%%
\section{The effective Hamiltonian}
\label{sec_hamiltonian}
%%%%%%%%%%%%%%%%%%%%%%%%%%%%%%%%%%%%%%%%%%%%%%%%%%%%%%%%%%%%%%%%%%%%%

In order to compute the NLO processes in the second-order quantum mechanical perturbation approach, we first formulate the QCD system of a heavy quarkonium and external light quarks and gluons in terms of an effective Hamiltonian. The non-relativistic heavy quarkonium system possesses a Hamiltonian
\begin{align}\label{H0}
H_0=\frac{{\vec p}^2}{m_Q}+V_1(r),
\end{align}
where $V_1$ is the binding potential for the color singlet $Q{\bar{Q}}$ (we ignore the weak repulsive potential in the color octet configuration). The coupling of the compact heavy quarkonium with external soft gluons can be summarized into an effective color-electric dipole type~\cite{Peskin:1979va,Yan:1980uh,Brambilla:2004jw,Voloshin:2007dx,Sumino:2014qpa} with two vertices
\begin{align}\label{H_QQbarg}
H_{Q\bar{Q}g}=V_{SO}+V_{OO},
\end{align}
where $V_{SO}$ represents the transition of a $Q\bar{Q}$ singlet state $|S\rangle$ to an octet $|O,a\rangle$ ($a=1,...,8$ denotes its color index) through interacting with an external gluon, which can be expressed in terms of the matrix element
\begin{align}\label{V_SO}
\langle O,a|V_{SO}|S \rangle &=\langle O,a|\frac{1}{2}g_s\vec{r}(\frac{\lambda^b}{2}-\frac{\bar{\lambda}^b}{2})\cdot\vec{E}^b|S\rangle \nonumber\\
&=\frac{g_s}{\sqrt{2N_c}}\vec{E}^a(t,\vec{x})\cdot \langle O|\vec{r}|S\rangle.
\end{align}
$V_{OO}$ in Eq.~(\ref{H_QQbarg}) represents the transition from an octet of color $b$ to another octet of color $a$ with the matrix element
\begin{align}\label{V_OO}
\langle O,a|V_{OO}|O,b \rangle =\frac{ig_s}{2}d^{abc}\vec{E}^c(t,\vec{x})\cdot \langle O|\vec{r}|O \rangle.
\end{align}
In Eqs.(\ref{V_SO}) and (\ref{V_OO}), $N_c=3$ is the number of colors in the fundamental representation of $SU(3)_c$, $\frac{\lambda^b}{2}$ ($\frac{\bar{\lambda}^b}{2}$) the color matrix of $Q$ ($\bar{Q}$), and $d^{abc}=2tr[\lambda^a/2\{\lambda^b/2,\lambda^c/2\}]$ a totally symmetric $SU(3)_c$ group invariant.

The Hamiltonian of the external light quarks, antiquarks and gluons reads~\cite{Peskin:1995}
\begin{align}\label{H_qqbarg}
H_{q\bar qg}&=\int d^3x {\psi^i}^{\dagger}(x)\Big[\beta m_q-i\vec{\alpha}\cdot\Big(\vec{\partial} -ig_s\vec{A}^a(\frac{\lambda^a}{2})^{ij}\Big)\Big]\psi^j(x) \nonumber\\ &+\frac{1}{2}\int d^3x\Big[{\vec{E^a}}^2(x)+\vec{B^a}^2(x)\Big],
\end{align}
where the $\vec{\alpha}$ and $\beta$ are Dirac matrices, $i,j=1,2,3$ the color indices of quarks, and $m_q$ the light quark mass. The self-interaction of gluons is implicit in the color-electric and color-magnetic field energy densities. We will work with the Weyl gauge $A_0^a(x)=0$~\cite{Sakurai-AQM:2008}, such that the color-electric field $\vec {E}^a(x)=\nabla A_0^a-\partial_0\vec A^a+g_sf^{abc}\vec A^bA_0^c$ appearing in Eqs.(\ref{V_SO}) and (\ref{V_OO}) reduces to
\begin{align}\label{color-electric_field}
	\vec {E}^a(t,\vec x)=-\frac{\partial}{\partial t}\vec{A}^a(t,\vec x).
\end{align}
Consequently, the self-interaction of gluons now arises only from the energy-density $1/2\vec B^a\cdot \vec B^a$ of the color-magnetic field $\vec B^a(x)=\nabla\times\vec A^a(x)-\frac{1}{2}g_sf^{abc}\vec A^b(x)\times \vec A^c(x)$. In particular, the three-gluon vertex (the four-gluon vertex appears at higher order in $g_s$ and is thus ignored here) reads
\begin{align}\label{V_3g}
V_{3g}&=-\frac{1}{4}g_sf^{abc}\int d^3x \Big[ (\nabla\times\vec A^a)\cdot(\vec A^b\times \vec A^c) \nonumber\\ &+(\vec A^b\times\vec A^c)\cdot(\nabla\times\vec A^a) \Big],
\end{align}
where $f^{abc}$ is the totally antisymmetric $SU(3)_c$ structure constant and the transverse gluon field is quantized in a box of volume $V$~\cite{Sakurai-AQM:2008}
\begin{align}\label{A_expansion}
\vec{A}^a(t,\vec x)=\sum_{\vec k, \lambda}\sqrt{\frac{1}{2V\omega_{ k}}}\vec{\epsilon}_{\vec{k},\lambda}\Big[a^a_{\vec{k}, \lambda}e^{i\vec{k}\cdot \vec{x}-i\omega_{k}t}+h.c.\Big],
\end{align}
with $\omega_{\vec k}$ being the gluon energy with momentum $\vec k$ and $\vec{\epsilon}_{\vec{k},\lambda}$ the polarization vector ($\lambda$=1, 2 denotes two transverse polarizations) that satisfies
\begin{align}\label{gluon_polarizations_sum}
\sum_{\lambda=1,2}\epsilon^{(i)}_{\vec{k}\lambda}\epsilon^{(j)}_{\vec{k}\lambda}=\delta^{ij}-\frac{k^{i}k^{j}}{\vec{k}^2}.
\end{align}
The creation and annihilation operators in Eq.(\ref{A_expansion}) satisfies the commutation relation $[a_{\vec k,\lambda}^a(t),{a_{\vec k',\lambda'}^{a'\dagger}}(t)]=\delta_{\vec k\vec k'}\delta_{\lambda\lambda'}\delta^{aa'}$, with one-gluon state $|g(\vec k,\lambda,a)\rangle=a_{\vec k,\lambda}^{a\dagger}|0\rangle$.

\begin{figure} [!t]
\includegraphics[width=1.0\columnwidth]{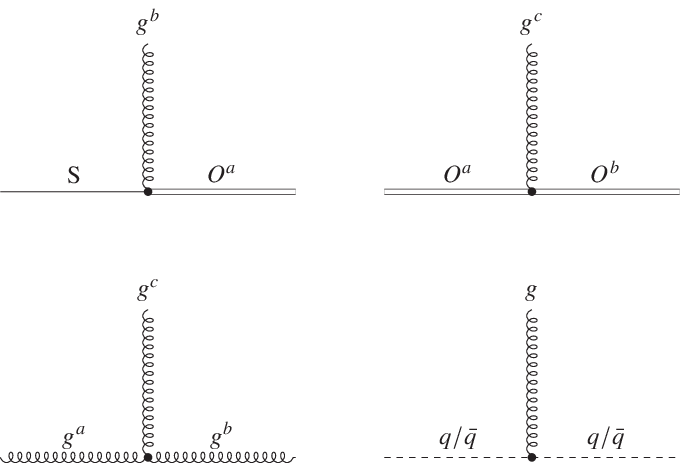}
\vspace{-0.3cm}
\caption{The four coupling vertices. The upper left, upper right, lower left and lower right refer to the $V_{SO}$, $V_{OO}$, $V_{3g}$ and $V_{q\bar{q}g}$ vertex, respectively. The single solid line, the double solid line, the single dashed line and the wavy line denote the heavy quarkonium singlet, the octet, the light quark/antiquark and the gluon, respectively.}
\label{Vertices}
\end{figure}

The coupling of the light quarks with gluons from Eq.({\ref{H_qqbarg}}) reads
\begin{align}\label{V_qqbarg}
V_{q\bar qg} = g_s \int d^3x {\psi^i}^{\dagger}(x)\vec{\alpha} \cdot \vec{A}^a(x)(\frac{\lambda^a}{2})^{ij}\psi^j(x).
\end{align}
The light quark fields are also quantized in a box of volume $V$~\cite{Sakurai-AQM:2008}
\begin{align}\label{q_expansion}
{\psi^j}(t,\vec x) = \sum_{\vec k,s}\sqrt{\frac{m_q}{E_{\vec k} V}} \Big(&b^j_{\vec k,s} u^{s}(\vec k)e^{i\vec k \cdot \vec x-iE_{\vec k}t} \nonumber\\	&+d^{j\dagger}_{\vec k,s} v^{s}(\vec k)e^{-i\vec k \cdot \vec x+iE_{\vec k}t}  \Big),
\end{align}	
\begin{align}\label{qdagger_expansion}
{\psi^i}^{\dagger}(t,\vec x)=\sum_{\vec k,s}\sqrt{\frac{m_q}{E_{\vec k} V}} \Big(&d^i_{\vec k,s} v^{s\dagger}(\vec k)e^{i\vec k \cdot \vec x-iE_{\vec k}t} \nonumber\\ &+b^{i\dagger}_{\vec k,s} u^{s\dagger}(\vec k)e^{-i\vec k \cdot \vec x+iE_{\vec k}t}  \Big),
\end{align}
where $E_{\vec{k}}=\sqrt{\vec{k}^2+m_q^2}$ is the energy of light quark (or antiquark) of momentum $\vec k$. The light quark/antiquark creation and annihilation operators satisfy the anti-commutation relation $\{b_{\vec k,s}^i(t),{b_{\vec k',s'}^{i'\dagger}}(t)\}=\{d_{\vec k,s}^i(t),{d_{\vec k',s'}^{i'\dagger}}(t)\}=\delta _{\vec k \vec k'}\delta_{s s'}\delta^{ii'}$, with one-quark state $|q(\vec{k},s,i) \rangle=b_{\vec{k} ,s}^{i\dagger}|0\rangle$ ($s,s'=1,2$ denotes the two spin states of the light quark or antiquark) and one-antiquark state $|\bar{q}(\vec{k},s,i)\rangle=d_{\vec{k},s}^{i\dagger}|0\rangle$. Following the conventions in~\cite{Sakurai-AQM:2008}, the orthogonal and normalization relation for spinors reads $u^{s\dagger}(\vec k)u^{s'}(\vec k)=v^{s\dagger}(\vec k)v^{s'}(\vec k)=\frac{E_{\vec{k}}}{m_q}\delta^{ss'}$, and the completeness relations are
\begin{align}\label{spinor_spin_sum}
\sum_{s=1}^{2}u_{\alpha}^{s}(\vec k)\bar u_{\beta}^{s}(\vec k)=\Big(\frac{-i\gamma \cdot k+m}{2m}\Big)_{\alpha \beta}, \\
\sum_{s=1}^{2}v_{\alpha}^{s}(\vec k)\bar v_{\beta}^{s}(\vec k)=-\Big(\frac{i\gamma \cdot k+m}{2m}\Big)_{\alpha \beta},
\end{align}
where $\bar u=u^\dagger \gamma_4$, $\bar v=v^\dagger \gamma_4$ and $\gamma \cdot k=\gamma_{\mu} k_{\mu}$ ($\,k_{\mu}=(\vec k, iE_{\vec k}))$ with the Hermitian matries $\gamma_{\mu}$ satisfying $\alpha_{k}=i\gamma_{4}\gamma_{k}$, $\gamma_4=\beta$ and $\{ \gamma_{\mu}, \, \gamma_{\nu}\}=2\delta_{\mu\nu}$.

To sum up, four coupling vertices come out of the QCD system of a heavy quarkonium and external light quarks and gluons, as depicted in Fig.~\ref{Vertices}, which will be used in Sec.~\ref{sec_dissociation} to construct the pertinent Feynman diagrams for the heavy quarkonium NLO dissociation processes.

%%%%%%%%%%%%%%%%%%%%%%%%%%%%%%%%%%%%%%%%%%%%%%%%%%%%%%%%%%%%%%%%
\section{Deriving the NLO dissociation cross sections}
\label{sec_dissociation}
%%%%%%%%%%%%%%%%%%%%%%%%%%%%%%%%%%%%%%%%%%%%%%%%%%%%%%%%%%%%%%%%

In this section we derive the NLO dissociation cross sections of heavy quarkonium due to inelastic scatterings by medium partons. We use the $s$-wave ground state charmonium $J/\psi$ as the example to illustrate the derivations, but in Sec.~\ref{sec_numerical_diss}, we will demonstrate numerical results for all heavy quarkonia including also the $p$-wave states. The derivations are performed within the approach of the second-order old-fashioned quantum mechanical perturbation theory~\cite{Sakurai-AQM:2008,Schwartz-QFT:2014,Greiner:1998,Gottfried-Yan:2003}, where all states are on-shell at all times and energy is not conserved at each vertex. The key quantity is the second-order transition amplitude
\begin{align}\label{2nd transition amplitude}
T_{fi}=\sum_m\frac{\langle f|V|m\rangle \langle m|V|i\rangle}{E_i-E_m+i\epsilon},
\end{align}
where the initial state $|i\rangle$, final state $|f\rangle$ and intermediate states $|m\rangle$ are all eigenstates of the {\it unperturbed} ({\it i.e.,} the free heavy quarkonium plus free light quarks/antiquarks and gluons system) Hamiltonian with eigenenergies $E_i=E_f$ and $E_m$, respectively, and $V$ denotes the coupling vertices (interaction Hamiltonian).

%%%%%%%%%%%%%%%%%%%%%%%%%%%%%%%%%%%%%%%%%%%%%%%%%%%%%%%%%%%%%%%%
\subsection{Dissociation involving $V_{SO}$ and $V_{OO}$}
\label{ssec_diss_a+b}
\begin{figure} [!t]
\includegraphics[width=1.0\columnwidth]{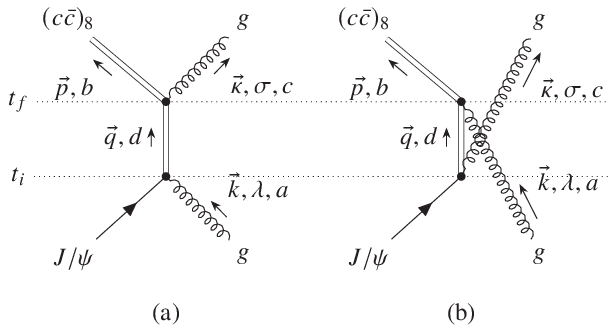}
\vspace{-0.3cm}
\caption{Feynman diagrams constructed from $V_{SO}$ and $V_{OO}$ for $g+J/\psi\to g+c+\bar{c}$. The time direction goes upward. Between the initial time $t_i$ and final time $t_f$ is the intermediate state.}
\label{figs[a]+[b]}
\end{figure}

%%%%%%%%%%%%%%%%%%%%%%%%%%%%%%%%%%%%%%%%%%%%%%%%%%%%%%%%%%%%%%%%
Using vertices $V_{SO}$ and $V_{OO}$, we can construct the Feynman diagrams (a) and (b) for the NLO dissociation of $J/\psi$ by an external gluon, $g+J/\psi\to g+c+\bar{c}$, as shown in Fig.~\ref{figs[a]+[b]}. These two diagrams are the counterparts of the $s$- and $u$-channel diagrams in relativistically covariant perturbation calculations for the photon-electron Compton scattering~\cite{{Peskin:1995}}. In the following, we work in the rest frame of the $J/\psi$ and neglect the three-momentum transfer from the incident gluon to the $J/\psi$. The latter approximation is justified by the fact that the mass of the heavy quarkonium is much larger than the typical momentum of thermal partons in the QGP. As a result, the center-of-mass momentum of the final state unbound octet $(c\bar{c})_8$ is also neglected and one deals only with the internal relative momentum of $(c\bar{c})_8$. The initial state for diagrams (a) and (b) consists of a $J/\psi$ at rest and an incident gluon of momentum $\vec k$, polarization $\lambda$ and color $a$, $|i\rangle=|J/\psi,g(\vec{k},\lambda,a)\rangle$, whereas the final state involves an unbound octet $(c\bar c)_8$ of internal relative momentum $\vec p$ and color $b$, plus an outgoing gluon of momentum $\vec\kappa$, polarization $\sigma$ and color $c$, $|f\rangle=|(c\bar c)_8(\vec{p}, b), g(\vec{\kappa},\sigma,c)\rangle$.

For diagram (a), the intermediate state is an octet of internal relative momentum $\vec q$ and color $d$, $|m\rangle=|(c\bar c)_8(\vec q,d)\rangle$. Then the transition matrix element due to $V_{SO}$ is calculated making use of Eqs.~({\ref{V_SO}), (\ref{color-electric_field}), (\ref{A_expansion})
\begin{align}\label{V_SO^a}
	&\langle m |V_{SO}| i \rangle = \langle (c\bar c)_8(\vec q,d) |V_{SO}| J/\psi, g(\vec{k},\lambda,a) \rangle\nonumber\\
	&=\frac{g_s}{\sqrt{6}}\langle 0|\vec{E}^d(t,\vec{x})|g(\vec{k},\lambda,a)\rangle \cdot\langle (c\bar c)_8(\vec{q})|\vec{r}|J/\psi\rangle \nonumber\\
	&=i\delta^{da}\frac{g_s}{\sqrt{6}}\sqrt{\frac{\omega_{\vec k}}{2V}}e^{i\vec{k}\cdot\vec{x}}\vec{\epsilon}_{\vec{k}\lambda}\cdot\langle (c\bar c)_8(\vec{q})|\vec{r}|J/\psi\rangle \nonumber\\
	&=g_s\frac{\delta^{da}}{V}\sqrt{\frac{\pi\omega_{\vec{k}}}{3}}e^{i\vec{k}\cdot\vec{x}}(\vec{\epsilon}_{\vec{k}\lambda}\cdot\vec{q})\frac{1}{q}\int r^3drj_1(qr)R_{10}(r),
\end{align}
where $R_{10}(r)$ is the normalized radial wave function for $J/\psi$. We have neglected the weak repulsive potential for the unbound octet $(c\bar c)_8$, so that the wave function for the internal relative motion is a plane wave $e^{i\vec{q}\cdot\vec{r}}/\sqrt{V}$. Upon a spherical wave expansion for the plane wave, $e^{i \vec{q} \cdot \vec{r} }= 4\pi\sum_{l,m}i^{l}j_{l}(qr)Y_{lm}(\theta,\phi)Y_{lm}^{*}(\theta',\phi')$ (primed angles for $\vec{q}$, and unprimed for $\vec{r}$), and use of the orthogonality relation for the spherical harmonics $ \int \int \sin\theta d\theta d\phi Y_{lm}^{*}(\theta,\phi)Y_{l'm'}(\theta,\phi)= \delta_{ll'} \delta_{mm'}$, only the first-order spherical Bessel function survives. Similarly, the other matrix element due to $V_{OO}$ is calculated using Eqs.~({\ref{V_OO}), (\ref{color-electric_field}), (\ref{A_expansion})
\begin{align}\label{V_OO^a}
	&\langle f |V_{OO}| m \rangle = \langle (c\bar c)_8(\vec{p},b), g(\vec{\kappa},\sigma,c) |V_{OO}| (c\bar c)_8(\vec q,d) \rangle \nonumber\\
	&=\frac{ig_s}{2}d^{bde} \langle  g(\vec{\kappa},\sigma,c) |\vec{E}^e(t,\vec{x})| 0 \rangle \cdot \langle (c\bar c)_8(\vec{p}) |\vec{r}| (c\bar c)_8(\vec{q}) \rangle \nonumber\\
	&=-d^{bdc}\frac{ig_s}{2V}(2\pi)^3\sqrt{\frac{\omega_{\vec{\kappa}}}{2V}}e^{-i\vec{\kappa}\cdot\vec{x}}\vec{\epsilon}_{\vec{\kappa}\sigma} \cdot \nabla_{\vec{q}}\delta^3(\vec{q}-\vec{p}),
\end{align}
where we have used $\int d^{3}r e^{i(\vec{q}-\vec{p}) \cdot \vec{r}} = (2\pi)^3 \delta^3(\vec{q}-\vec{p})$  for the continuous momenta.

Now substituting these two matrix elements into Eq.~(\ref{2nd transition amplitude}) and carrying out the summation over the momentum and color of the intermediate state $\sum_{m}=\sum_{\vec{q}}\sum_{d}=V/(2\pi)^3\int d^{3}\vec q \sum_{d}$, one arrives at the transition amplitude for diagram (a) after an integration by part to integrate out the $\delta^3(\vec{q}-\vec{p})$
\begin{align}\label{T_fi^a} T_{fi}^{(a)}=&d^{abc}\frac{ig_s^2}{2V}\sqrt{\frac{\pi\omega_{\vec{k}}\omega_{\vec{\kappa}}}{6V}}e^{i\vec{k}\cdot\vec{x}-i\vec{\kappa}\cdot\vec{x}} \nonumber\\ 	&\times\vec{\epsilon}_{\vec{\kappa}\sigma}\cdot\Big[A(p,k)\vec{\epsilon}_{\vec{k}\lambda}+(\vec{\epsilon}_{\vec{k}\lambda}\cdot\vec{p})\frac{\vec{p}}{p^2}B(p,k)\Big],
\end{align}
where
\begin{align}\label{AB(p,k)}
	&A(p,k)=\frac{\int r^3drj_1(pr)R_{10}(r) }{p(-\epsilon_B+\omega_{\vec{k}}-\frac{p^2}{m_Q}+i\epsilon)},\nonumber\\
	&B(p,k)=\frac{\frac{2p}{m_Q}\int r^3drj_1(pr)R_{10}(r) }{(-\epsilon_B+\omega_{\vec{k}}-\frac{p^2}{m_Q}+i\epsilon)^2}-\frac{\int r^4drj_2(pr)R_{10}(r) }{(-\epsilon_B+\omega_{\vec{k}}-\frac{p^2}{m_Q}+i\epsilon)}.
\end{align}
contain the bound state wave function and binding energy $\epsilon_B$. Here all energies are measured relative to the threshold of the $c\bar{c}$ pair.

For diagram (b), the intermediate $|m \rangle=| (c\bar c)_8(\vec q,d),\,g(\vec{k},\lambda,a),\, g(\vec{\kappa},\sigma,c) \rangle$ involves two gluons that have the same quantum numbers as the incident and outgoing gluons, respectively. Similar manipulations as above lead to the matrix elements due to $V_{SO}$ and $V_{OO}$, respectively
\begin{align}\label{V_SO^b}
	&\langle m |V_{SO}| i \rangle = \langle (c\bar c)_8(\vec q,d),  g(\vec{\kappa},\sigma,c) |V_{SO}| J/\psi \rangle\nonumber\\ &=-g_s\frac{\delta^{dc}}{V}\sqrt{\frac{\pi\omega_{\vec{\kappa}}}{3}}e^{-i\vec{\kappa}\cdot\vec{x}}(\vec{\epsilon}_{\vec{\kappa}\sigma}\cdot\vec{q})\frac{1}{q}\int r^3drj_1(qr)R_{10}(r),
\end{align}
\begin{align}\label{V_OO^b}
	&\langle f |V_{OO}| m \rangle = \langle (c\bar c)_8(\vec{p},b) |V_{OO}| (c\bar c)_8(\vec q,d), g(\vec{k},\lambda,a) \rangle \nonumber\\
	&=d^{bda}\frac{ig_s}{2V}(2\pi)^3\sqrt{\frac{\omega_{\vec{k}}}{2V}}e^{i\vec{k}\cdot\vec{x}}\vec{\epsilon}_{\vec{k}\lambda} \cdot \nabla_{\vec{q}}\delta^3(\vec{q}-\vec{p}).
\end{align}
Finally the transition amplitude for diagram (b) reads
\begin{align}\label{T_fi^b} T_{fi}^{(b)}=&d^{abc}\frac{ig_s^2}{2V}\sqrt{\frac{\pi\omega_{\vec{k}}\omega_{\vec{\kappa}}}{6V}}e^{i\vec{k}\cdot\vec{x}-i\vec{\kappa}\cdot\vec{x}}\nonumber\\ 	&\times\vec{\epsilon}_{\vec{k}\lambda}\cdot\Big[C(p,\kappa)\vec{\epsilon}_{\vec{\kappa}\sigma}+(\vec{\epsilon}_{\vec{\kappa}\sigma}\cdot\vec{p})\frac{\vec{p}}{p^2}D(p,\kappa)\Big],
\end{align}
where
\begin{align}\label{CD(p,kapp)}
	&C(p,\kappa)=\frac{\int r^3drj_1(pr)R_{10}(r)} {p(-\epsilon_B-\omega_{\vec{\kappa}}-\frac{p^2}{m_Q}+i\epsilon)},\nonumber\\
	&D(p,\kappa)=\frac{\frac{2p}{m_Q}\int r^3drj_1(pr)R_{10}(r)} {(-\epsilon_B-\omega_{\vec{\kappa}}-\frac{p^2}{m_Q}+i\epsilon)^2}-\frac{\int r^4drj_2(pr)R_{10}(r) }{(-\epsilon_B-\omega_{\vec{\kappa}}-\frac{p^2}{m_Q}+i\epsilon)}.
\end{align}
again contain the information of bound state wave function and binding energy.

%%%%%%%%%%%%%%%%%%%%%%%%%%%%%%%%%%%%%%%%%%%%%%%%%%%%%%%%%%%%%%%%
\subsection{Dissociation involving $V_{SO}$ and $V_{3g}$}
\label{ssec_diss_c+d}
%%%%%%%%%%%%%%%%%%%%%%%%%%%%%%%%%%%%%%%%%%%%%%%%%%%%%%%%%%%%%%%%
The NLO dissociation of $J/\psi$ by an external gluon, $g+J/\psi\to g+c+\bar{c}$ can be also constructed using the vertices of $V_{SO}$ and $V_{3g}$, as represented by the Feynman diagrams (c) and (d) in Fig.~\ref{figs[c]+[d]}. These two diagrams have the same initial and final state as specified for diagrams (a) and (b).

\begin{figure} [!t]
\includegraphics[width=1.0\columnwidth]{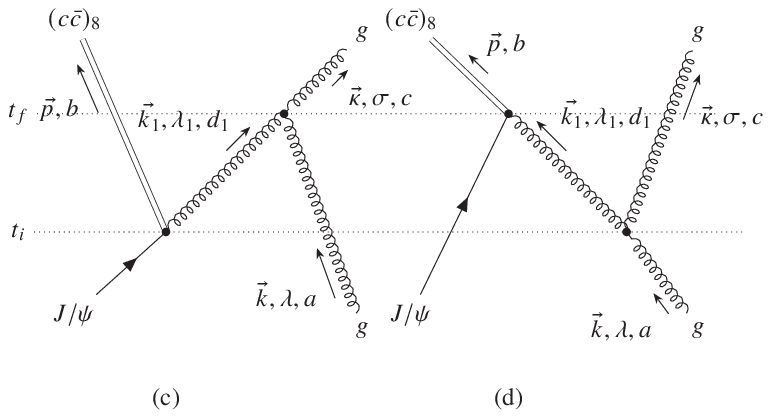}
\vspace{-0.3cm}
\caption{Feynman diagrams constructed from $V_{SO}$ and $V_{3g}$ for $g+J/\psi\to g+c+\bar{c}$. The time direction goes upward. Between the initial time $t_i$ and final time $t_f$ is the intermediate state.}
\label{figs[c]+[d]}
\end{figure}

For diagram (c), the intermediate state contains an octet with the same quantum numbers as the $(c\bar{c})_8$ in the final state, a gluon with the same quantum numbers as the incident gluon, and another gluon whose momentum, polarization and color are not fixed and thus denoted as $(\vec{k_1}$, $\lambda_1$ and $d_1)$, respectively, such that $|m\rangle = |(c\bar c)_8(\vec p,b), g(\vec{k},\lambda,a), g(\vec{k_1},\lambda_1,d_1) \rangle$. Similar to Eq.~(\ref{V_SO^b}), the transition matrix element due to $V_{SO}$ reads
\begin{align}\label{V_SO^c}
& \langle m |V_{SO}| i \rangle = \langle (c\bar c)_8(\vec{p},b),\,g(\vec{k}_1,\lambda_1,d_1) |V_{SO}| J/\psi \rangle\nonumber\\	&=-g_s\frac{\delta^{d_{1}b}}{V}\sqrt{\frac{\pi\omega_{\vec{k}_1}}{3}}e^{-i\vec{k}_1\cdot\vec{x}}(\vec{\epsilon}_{\vec{k}_1\lambda_1}\cdot\vec{p})\frac{1}{p}\int r^3drj_1(pr)R_{10}(r).
\end{align}
To calculate the transition matrix element due to $V_{3g}$, we substitute Eq.~(\ref{A_expansion}) into Eq.~(\ref{V_3g}) and obtain the relevant term involving the combination of two annihilation operators and one creation operator
\begin{align}\label{V_{3g}a+aa} &V_{3g}(a^{\dagger}aa)=f^{abc}\frac{ig_s}{4}\sum_{\vec{k}_1\lambda_1}\sum_{\vec{k}_2\lambda_2}\sum_{\vec{k}_3\lambda_3}\sqrt{\frac{1}{2V\omega_{\vec{k}_1}\omega_{\vec{k}_2}\omega_{\vec{k}_3}}}\nonumber\\ 	&(\vec{\epsilon}_{\vec{k}_1\lambda_1}\times\vec{k}_1)\cdot(\vec{\epsilon}_{\vec{k}_2\lambda_2}\times\vec{\epsilon}_{\vec{k}_3\lambda_3})  \Big(a_{\vec k_2\lambda_2}^{b\dagger}a_{\vec k_3\lambda_3}^{c}a_{\vec k_1\lambda_1}^{a}\delta_{\vec{k}_2, \vec{k}_3+\vec{k}_1}\nonumber\\ 	&+a_{\vec k_3\lambda_3}^{c\dagger}a_{\vec k_1\lambda_1}^{a}a_{\vec k_2\lambda_2}^{b}\delta_{\vec{k}_3, \vec{k}_1+\vec{k}_2}	-a_{\vec k_1\lambda_1}^{a\dagger}a_{\vec k_2\lambda_2}^{b}a_{\vec k_3\lambda_3}^{c}\delta_{\vec{k}_1, \vec{k}_2+\vec{k}_3}\Big),
\end{align}
which yields
\begin{align}\label{V_3g_c}
	& \langle f |V_{3g}| m \rangle = \langle g(\vec{\kappa},\sigma,c) |V_{3g}(a^{\dagger}aa)| g(\vec{k},\lambda,a),\, g(\vec{k}_1,\lambda_1,d_1) \rangle\nonumber\\
	&=f^{abc}\frac{ig_s}{2}\sqrt{\frac{1}{2V\omega_{\vec{k}}\omega_{\vec{\kappa}}\omega_{\vec{k}_1}}} \delta_{\vec{\kappa}, \vec{k}+\vec{k}_1}\Big[ (\vec{\epsilon}_{\vec{k}\lambda}\times\vec{k}) \cdot (\vec{\epsilon}_{\vec{k}_1\lambda_1} \times\vec{\epsilon}_{\vec{\kappa}\sigma})\nonumber\\ &-(\vec{\epsilon}_{\vec{k}_1\lambda_1}\times\vec{k}_1) \cdot (\vec{\epsilon}_{\vec{k}\lambda}\times\vec{\epsilon}_{\vec{\kappa}\sigma}) +(\vec{\epsilon}_{\vec{\kappa}\sigma}\times\vec{\kappa}) \cdot (\vec{\epsilon}_{\vec{k}_1\lambda_1}\times\vec{\epsilon}_{\vec{k}\lambda})\Big] \nonumber\\ &=f^{abc}\frac{ig_s}{2}\sqrt{\frac{1}{2V\omega_{\vec{k}}\omega_{\vec{\kappa}}\omega_{\vec{k}_1}}}\vec{\epsilon}_{\vec{k}_1\lambda_1}\cdot\Big[2\vec{\epsilon}_{\vec{k}\lambda}(\vec{k}\cdot\vec{\epsilon}_{\vec{\kappa}\sigma})\nonumber\\ &+2\vec{\epsilon}_{\vec{\kappa}\sigma}(\vec{\kappa}\cdot\vec{\epsilon}_{\vec{k}\lambda})-(\vec{k}+\vec{\kappa})(\vec{\epsilon}_{\vec{k}\lambda}\cdot\vec{\epsilon}_{\vec{\kappa}\sigma})\Big]\delta_{\vec{\kappa}, \vec{k}+\vec{k}_1}.
\end{align}

Substituting these two matrix elements into Eq.~(\ref{2nd transition amplitude}) and performing the summation over the intermediate state $\sum_{m} = \sum_{\vec{k}_1}\sum_{\lambda_1}\sum_{d_1}$, we obtain the transition amplitude for diagram (c)
\begin{align}\label{T_fi^c}
	T_{fi}^{(c)}&=f^{abc}\frac{-ig_s^2}{V}\sqrt{\frac{\pi}{6V\omega_{\vec{k}}\omega_{\vec{\kappa}}}}e^{-i(\vec{\kappa}-\vec{k})\cdot\vec{x}} \nonumber\\ 	&\times \Big\{(\vec{p}\cdot\vec{\epsilon}_{\vec{k}\lambda}) (\vec{k}\cdot \vec{\epsilon}_{\vec{\kappa}\sigma}) + (\vec{p}\cdot\vec{\epsilon}_{\vec{\kappa}\sigma}) (\vec{\kappa}\cdot\vec{\epsilon}_{\vec{k}\lambda}) \nonumber\\ 	&-\frac{(\vec{\epsilon}_{\vec{k}\lambda}\cdot\vec{\epsilon}_{\vec{\kappa}\sigma})} {(\vec{\kappa}-\vec{k})^2}\Big[(\vec{p}\cdot\vec{k})(\vec{\kappa}^2-\vec{\kappa}\cdot\vec{k}) +(\vec{p}\cdot\vec{\kappa})(\vec{k}^2-\vec{\kappa}\cdot\vec{k})\Big] \Big\} \nonumber\\		&\times\frac{\int r^3drj_1(pr)R_{10}(r)} {p(-\epsilon_B-\frac{p^2}{m_Q}-\omega_{(\vec{\kappa}-\vec{k})}+i\epsilon)}.
\end{align}
where $\omega_{(\vec{\kappa}-\vec{k})}$ is the energy of the intermediate gluon with $\vec{k}_1=\vec\kappa-\vec{k}$. To arrive at Eq.~(\ref{T_fi^c}), the completeness relation for the gluon polarization vector Eq.~(\ref{gluon_polarizations_sum}) has been used for $\vec\epsilon_{\vec k_1\lambda_1}$.

For diagram (d), the intermediate state $| m \rangle = | J/\psi, g(\vec{\kappa},\sigma,c), g(\vec{k_1},\lambda_1,d_1) \rangle$. The term proportional to the combination of one annihilation operator and two creation operators in the expansion of the $V_{3g}$ upon using Eq.~(\ref{A_expansion}), {\it i.e.},
\begin{align}\label{V_{3g}a+a+a} &V_{3g}(a^{\dagger}a^{\dagger}a)=f^{abc}\frac{ig_s}{4}\sum_{\vec{k}_1,\lambda_1}\sum_{\vec{k}_2,\lambda_2}\sum_{\vec{k}_3,\lambda_3}\sqrt{\frac{1}{2V\omega_{\vec{k}_1}\omega_{\vec{k}_2}\omega_{\vec{k}_3}}}\nonumber\\ 	&(\vec{\epsilon}_{\vec{k}_1\lambda_1}\times\vec{k}_1)\cdot(\vec{\epsilon}_{\vec{k}_2\lambda_2}\times\vec{\epsilon}_{\vec{k}_3\lambda_3}) \Big(-a_{\vec k_3\lambda_3}^{c\dagger}a_{\vec k_1\lambda_1}^{a\dagger}a_{\vec k_2\lambda_2}^{b}\delta_{\vec{k}_2, \vec{k}_3+\vec{k}_1} \nonumber\\ 	&-a_{\vec k_1\lambda_1}^{a\dagger}a_{\vec k_2\lambda_2}^{b\dagger}a_{\vec k_3\lambda_3}^{c}\delta_{\vec{k}_3, \vec{k}_1+\vec{k}_2} +a_{\vec k_2\lambda_2}^{b\dagger}a_{\vec k_3\lambda_3}^{c\dagger}a_{\vec k_1\lambda_1}^{a}\delta_{\vec{k}_1, \vec{k}_2+\vec{k}_3}\Big),
\end{align}
is responsible for the transition matrix element due to $V_{3g}$, which reads
\begin{align}\label{V_3g_d}
	& \langle m |V_{3g}| i \rangle = \langle g(\vec{\kappa},\sigma,c), g(\vec{k_1},\lambda_1,d_1) | V_{3g}(a^{\dagger}a^{\dagger}a)| g(\vec{k},\lambda,a) \rangle\nonumber\\
	&=f^{abc}\frac{ig_s}{2}\sqrt{\frac{1}{2V\omega_{\vec{k}}\omega_{\vec{\kappa}}\omega_{\vec{k}_1}}}\delta_{\vec{k}, \vec{k}_1+\vec{\kappa}} \Big[ (\vec{\epsilon}_{\vec{k}\lambda}\times\vec{k}) \cdot (\vec{\epsilon}_{\vec{k}_1\lambda_1}\times\vec{\epsilon}_{\vec{\kappa}\sigma})\nonumber\\ &+(\vec{\epsilon}_{\vec{k}_1\lambda_1}\times\vec{k}_1) \cdot (\vec{\epsilon}_{\vec{k}\lambda}\times\vec{\epsilon}_{\vec{\kappa}\sigma}) +(\vec{\epsilon}_{\vec{\kappa}\sigma}\times\vec{\kappa}) \cdot (\vec{\epsilon}_{\vec{k}_1\lambda_1}\times\vec{\epsilon}_{\vec{k}\lambda})\Big] \nonumber\\
	&=f^{abc}\frac{ig_s}{2}\sqrt{\frac{1}{2V\omega_{\vec{k}}\omega_{\vec{\kappa}}\omega_{\vec{k}_1}}}\vec{\epsilon}_{\vec{k}_1\lambda_1} \cdot\Big[2\vec{\epsilon}_{\vec{k}\lambda}(\vec{k}\cdot\vec{\epsilon}_{\vec{\kappa}\sigma})\nonumber\\ 	&+2\vec{\epsilon}_{\vec{\kappa}\sigma}(\vec{\kappa}\cdot\vec{\epsilon}_{\vec{k}\lambda})-(\vec{k}+\vec{\kappa})(\vec{\epsilon}_{\vec{k}\lambda}\cdot\vec{\epsilon}_{\vec{\kappa}\sigma})\Big]\delta_{\vec{k},\vec{k}_1+\vec{\kappa}}.
\end{align}
The other transition matrix element due to $V_{SO}$ is
\begin{align}\label{V_SO^d}
& \langle f |V_{SO}| m \rangle = \langle (c\bar c)_8(\vec{p},b)|V_{SO}| J/\psi,\,  g(\vec{k}_1,\lambda_1,d_1)  \rangle\nonumber\\
	&=g_s\frac{\delta^{d_{1}b}}{V}\sqrt{\frac{\pi\omega_{\vec{k}_1}}{3}}e^{i\vec{k}_1\cdot\vec{x}}(\vec{\epsilon}_{\vec{k}_1\lambda_1}\cdot\vec{p})\frac{1}{p}\int r^3drj_1(pr)R_{10}(r),
\end{align}

Similar treatments as taken to arrive at Eq.~(\ref{T_fi^c}) lead to the transition amplitude for diagram (d)
\begin{align}\label{T_fi^d}
	T_{fi}^{(d)}&=f^{abc}\frac{ig_s^2}{V}\sqrt{\frac{\pi}{6V\omega_{\vec{k}}\omega_{\vec{\kappa}}}}e^{i(\vec{k}-\vec{\kappa})\cdot\vec{x}}\nonumber\\
	&\times\Big\{(\vec{p}\cdot\vec{\epsilon}_{\vec{k}\lambda})(\vec{k}\cdot\vec{\epsilon}_{\vec{\kappa}\sigma}) + (\vec{p}\cdot\vec{\epsilon}_{\vec{\kappa}\sigma})(\vec{\kappa}\cdot\vec{\epsilon}_{\vec{k}\lambda}) \nonumber\\
	&-\frac{(\vec{\epsilon}_{\vec{k}\lambda}\cdot\vec{\epsilon}_{\vec{\kappa}\sigma})} {(\vec{\kappa}-\vec{k})^2}\Big[(\vec{p}\cdot\vec{k})(\vec{\kappa}^2-\vec{\kappa}\cdot\vec{k}) +(\vec{p}\cdot\vec{\kappa})(\vec{k}^2-\vec{\kappa}\cdot\vec{k})\Big] \Big\}\nonumber\\	&\times\frac{\int r^3drj_1(pr)R_{10}(r)} {p(\omega_{\vec{k}}-\omega_{\vec{\kappa}}-\omega_{(\vec{\kappa}-\vec{k})}+i\epsilon)},
\end{align}
which is of opposite sign compared to $T_{fi}^{(c)}$ in Eq.~(\ref{T_fi^c}).

%%%%%%%%%%%%%%%%%%%%%%%%%%%%%%%%%%%%%%%%%%%%%%%%%%%%%%%%%%%%%%%%
\subsection{Dissociation involving $V_{SO}$ and $V_{q\bar qg}$}
\label{ssec_diss_e+f}
%%%%%%%%%%%%%%%%%%%%%%%%%%%%%%%%%%%%%%%%%%%%%%%%%%%%%%%%%%%%%%%%
The Feynman diagrams for NLO dissociation of heavy quarkonium by thermal quarks/antiquarks, $q/\bar{q}+J/\psi\to q/\bar{q}+c+\bar{c}$, as shown in Fig.~\ref{figs[e]+[f]}, are constructed using the vertices $V_{SO}$ and $V_{q\bar qg}$. Here the initial state $|i\rangle=|J/\psi, q/\bar{q}(\vec{k},r,i) \rangle$ and the final state $|f\rangle=|(c\bar c)_8(\vec{p}, b), q/\bar{q}(\vec{\kappa},s,j)\rangle$, with $i,j=1,2,3$ labeling the quark/antiquark colors and $r,s=1,2$ their spins. In the following, we focus on the light quark of a given flavor and the terms in $V_{q\bar qg}$ (upon using expansions  Eqs.~(\ref{A_expansion}), (\ref{q_expansion}), (\ref{qdagger_expansion})) that are relevant for diagram (e) and (f) are

\begin{figure} [!t]
	\includegraphics[width=1.0\columnwidth]{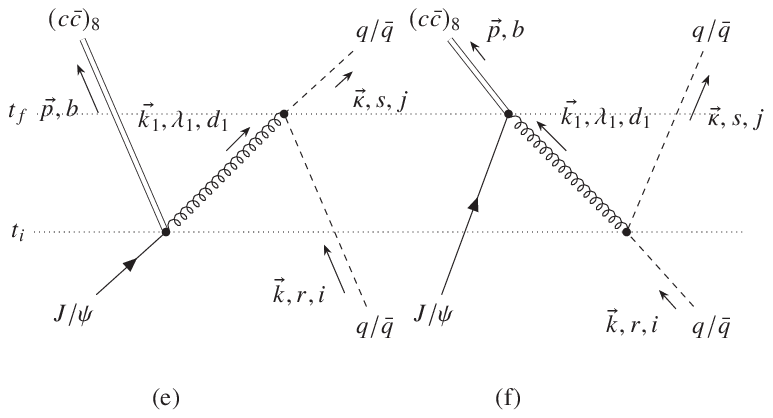}
	\vspace{-0.3cm}
	\caption{Feynman diagrams constructed from $V_{SO}$ and $V_{q\bar qg}$ for $q/\bar{q}+J/\psi\to q/\bar{q}+c+\bar{c}$. The time direction goes upward. Between the initial time $t_i$ and final time $t_f$ is the intermediate state.}
	\label{figs[e]+[f]}
\end{figure}

\begin{align}\label{v_b+ba}
	&V_{q\bar{q}g}(b^{\dagger}ba)=g_sm_q\sum_{\vec k, \lambda}\sum_{\vec{k}_1,s_1} \sum_{\vec{k}_2,s_2}\sqrt{\frac{1}{2V\omega_{\vec{k}}E_{\vec{k}_1}E_{\vec{k}_2}}}\delta_{\vec{k}_2, \vec{k}+\vec{k}_1} \nonumber\\ 	&\times  b_{\vec{k}_2,s_2}^{i_2\dagger}(t)u^{s_2\dagger}(\vec{k}_2)\vec{\alpha}\cdot \vec{\epsilon}_{\vec{k}\lambda}a_{\vec{k}\lambda}^{a}(t)\Big(\frac{\lambda^a}{2}\Big)^{i_2i_1}b_{\vec{k}_1,s_1}^{i_1}(t)u^{s_1}(\vec{k}_1),
\end{align}
\begin{align}\label{v_b+ba+}
	&V_{q\bar{q}g}(b^{\dagger}ba^{\dagger})=g_sm_q\sum_{\vec k, \lambda}\sum_{\vec{k}_1,s_1} \sum_{\vec{k}_2,s_2}\sqrt{\frac{1}{2V\omega_{\vec{k}}E_{\vec{k}_1}E_{\vec{k}_2}}}\delta_{\vec{k}_1, \vec{k}+\vec{k}_2} \nonumber\\
	&\times b_{\vec{k}_2,s_2}^{i_2\dagger}(t)u^{s_2\dagger}(\vec{k}_2)\vec{\alpha}\cdot \vec{\epsilon}_{\vec{k}\lambda}a_{\vec{k}\lambda}^{a\dagger}(t)\Big(\frac{\lambda^a}{2}\Big)^{i_2i_1}b_{\vec{k}_1,s_1}^{i_1}(t)u^{s_1}(\vec{k}_1).
\end{align}

For diagram (e), the intermediate state $ | m \rangle = | (c\bar c)_8(\vec p,b), q(\vec{k},r,i), g(\vec{k_1},\lambda_1,d_1) \rangle$. The transition matrix element $\langle m |V_{SO}| i \rangle$ is same as Eq.~(\ref{V_SO^c}). The other transition matrix element due to $V_{q\bar{q}g}$ reads
\begin{align}
	&\langle f |V_{q\bar{q}g}| m \rangle = \langle q(\vec{\kappa},s,j) | V_{q\bar{q}g}(b^{\dagger}ba) | g(\vec{k}_1,\lambda_1,d_1), q(\vec{k},r,i) \rangle \nonumber\\
	&=g_s m_q \sqrt{\frac{1}{2V\omega_{\vec{k}_1}E_{\vec{k}}E_{\vec{\kappa}}}}  u^{s\dagger}(\vec{\kappa}) \vec{\alpha} \cdot \vec{\epsilon}_{\vec{k}_1\lambda_1} \Big(\frac{\lambda^{d_1}}{2}\Big)^{ji} u^{r}(\vec{k})\delta_{\vec{\kappa}, \vec{k}+\vec{k}_1}.
\end{align}
The transition amplitude for diagram (e) is then computed from Eq.~(\ref{2nd transition amplitude}) with $\sum_{m} = \sum_{\vec{k}_1}\sum_{\lambda_1}\sum_{d_1}$:
\begin{align}\label{T_fi^e}
	T_{fi}^{(e)}&=\sum_{m}\frac{\langle f |V_{q\bar{q}g}| m \rangle \langle m |V_{SO}| i  \rangle}{E_i-E_m+i\epsilon}\nonumber\\
	&=-\frac{g_s^2m_q}{V}\sqrt{\frac{\pi}{6VE_{\vec{k}}E_{\vec{\kappa}}}}e^{-i(\vec{\kappa}-\vec{k})\cdot\vec{x}}\Big[ u^{s\dagger}(\vec{\kappa})\vec{\alpha}\cdot\vec{p}u^{r} (\vec{k}) \nonumber\\ 	&-\frac{[u^{s\dagger}(\vec{\kappa})(\vec{\kappa}-\vec{k})\cdot\vec{\alpha}u^{r}(\vec{k})][(\vec{\kappa}-\vec{k})\cdot\vec{p}]}{|\vec{\kappa}-\vec{k}|^2} \Big] \Big(\frac{\lambda^b}{2}\Big)^{ji}\nonumber\\ 	&\times \frac{\int r^3drj_1(pr)R_{10}(r)}  {p(-\epsilon_B-\frac{p^2}{m_Q}-\omega_{(\vec{\kappa}-\vec{k})}+i\epsilon)}.
\end{align}
To arrive at Eq.~(\ref{T_fi^e}), the completeness relation for the gluon polarization vector Eq.~(\ref{gluon_polarizations_sum}) has been again used for $\vec\epsilon_{\vec k_1\lambda_1}$.

Finally for diagram (f), the intermediate state $| m \rangle = | J/\psi , q(\vec{\kappa},s,j), g(\vec{k_1},\lambda_1,d_1) \rangle$. The transition matrix element due to $V_{q\bar{q}g}$ reads
\begin{align}
	& \langle m |V_{q\bar{q}g}| i \rangle = \langle q(\vec{\kappa},s,j),\, g(\vec{k_1},\lambda_1,d_1) |V_{q\bar{q}g}(b^{\dagger}ba^{\dagger})| q(\vec{k},r,i) \rangle \nonumber\\
	&=g_sm_q\sqrt{\frac{1}{2V\omega_{\vec{k}_1}E_{\vec{k}}E_{\vec{\kappa}}}}u^{s\dagger}(\vec{\kappa})\vec{\alpha}\cdot\vec{\epsilon}_{\vec{k}_1\lambda_1}\Big(\frac{\lambda^{d_1}}{2}\Big)^{ji}u^{r}(\vec{k})\delta_{\vec{k}, \vec{k}_1+\vec{\kappa}}.
\end{align}
The other matrix element due to $V_{SO}$ ($\langle f |V_{SO}| m \rangle$) is same as Eq.~(\ref{V_SO^d}). The transition amplitude for diagram (f) is obtained by combining these two matrix elements into Eq.~(\ref{2nd transition amplitude}) and carrying out the summation over intermediate states $\sum_{m} = \sum_{\vec{k}_1}\sum_{\lambda_1}\sum_{d_1}$, such that
\begin{align}\label{T_fi^f}
	T_{fi}^{(f)}&=+\frac{g_s^2m_q}{V}\sqrt{\frac{\pi}{6VE_{\vec{k}}E_{\vec{\kappa}}}}e^{i(\vec{k}-\vec{\kappa})\cdot\vec{x}}\Big[ u^{s\dagger}(\vec{\kappa})\vec{\alpha}\cdot\vec{p}u^{r} (\vec{k})\nonumber\\  	&-\frac{[u^{s\dagger}(\vec{\kappa})(\vec{k}-\vec{\kappa})\cdot\vec{\alpha}u^{r}(\vec{k})][(\vec{k}-\vec{\kappa})\cdot\vec{p}]}{|\vec{k}-\vec{\kappa}|^2} \Big] \Big(\frac{\lambda^b}{2}\Big)^{ji}\nonumber\\  	&\times\frac{\int r^3drj_1(pr)R_{10}(r)} {p(E_{\vec{k}}-E_{ \vec{\kappa}}-\omega_{(\vec{\kappa}-\vec{k})}+i\epsilon)},
\end{align}
which is of opposite sign relative to $T_{fi}^{(e)}$ in Eq.~(\ref{T_fi^e}).

The transition amplitudes for $\bar{q}+J/\psi\to \bar{q}+c+\bar{c}$ are obtained upon replacement of spinors in Eqs.~(\ref{T_fi^e}) and (\ref{T_fi^f}) via $u_s^{\dagger}(\vec \kappa)\rightarrow v_r^{\dagger}(\vec k)$, $u_r(\vec k)\rightarrow v_s(\vec \kappa)$.

%%%%%%%%%%%%%%%%%%%%%%%%%%%%%%%%%%%%%%%%%%%%%%%%%%%%%%%%%%%%%%%%
\subsection{Dissociation cross sections}
\label{ssec_diss_cross-sections}
%%%%%%%%%%%%%%%%%%%%%%%%%%%%%%%%%%%%%%%%%%%%%%%%%%%%%%%%%%%%%%%%
Using the second-order transition amplitudes calculated above, the transition rates for the NLO dissociation processes are calculated from
Fermi's golden rule~\cite{Sakurai-AQM:2008,Greiner:1998,Gottfried-Yan:2003}
\begin{align}\label{fermi's_golden_rule}
W_{i\to f}=2\pi|T_{fi}|^2\delta(E_i-E_f).
\end{align}
The cross section is obtained by dividing the transition rate by the flux of incident parton $v_{rel}/V$ ($v_{rel}$ being the relative velocity between the incident parton and the $J/\psi$ target) and averaging (summing) over the initial (final) state degeneracies.

We note that both $T_{fi}^{(a)}$, $T_{fi}^{(b)}\propto d^{abc}$, whereas both $T_{fi}^{(c)}$, $T_{fi}^{(d)}\propto f^{abc}$. The totally symmetric and totally antisymmetry property of $d^{abc}$ and $f^{abc}$, respectively, implies $\sum_{abc}d^{abc}f^{abc}=0$, so that there's no interference between $T_{fi}^{(a)}+T_{fi}^{(b)}$ and $T_{fi}^{(c)}+T_{fi}^{(d)}$ and they have independent cross sections. For diagrams (a) and (b), the summation and averaging procedure gives the cross section
\begin{align}\label{sigma_ab}
	&\sigma^{(a+b)}(E_g)= 2\pi V \frac{V}{(2\pi)^3}\int d^3p\sum_{b} \frac{V}{(2\pi)^3} \int d^3\kappa\sum_{\sigma}\sum_{c}\nonumber\\  &\times \frac{1}{4\pi}\int d\Omega_{\vec{k}}\frac{1}{2}\sum_{\lambda}\frac{1}{8}\sum_{a} |T_{fi}^{(a)}+T_{fi}^{(b)}|^2\delta(E_i-E_f)\nonumber\\
	&=\frac{5}{216\pi^2}g_s^4E_{g}\int p^{2}dp\int \kappa^2\omega_{\vec{\kappa}}d\kappa\{\cdots\} \nonumber\\	&\times\delta(-\epsilon_B+\omega_{\vec{k}}-\frac{p^2}{m_Q}-\omega_{\vec{\kappa}}),
\end{align}
where $E_g=\omega_{\vec k}$ is the incident gluon energy and
\begin{align}
	\{\cdots\}=&A^2(p,k)+C^2(p,\kappa)+2A(p,k)C(p,\kappa)\nonumber\\	&+\frac{1}{3}\Big\{B^2(p,k)+D^2(p,\kappa)+2\Big[A(p,k)B(p,k)\nonumber\\	&+A(p,k)D(p,\kappa)+B(p,k)C(p,\kappa)\nonumber\\	&+B(p,k)D(p,\kappa)+C(p,\kappa)D(p,\kappa)\Big]\Big\}.
\end{align}
To arrive at second equality of Eq.~(\ref{sigma_ab}), two identities regarding the totally symmetric $SU(3)_c$ group invariant $\sum_{abc}d^{abc}d^{abc}=40/3$ and the polarization vector ($\vec\rho$ being an arbitrary vector independent of $\Omega_{\vec k}$)
\begin{align}
\frac{1}{4\pi}\int d\Omega_{\vec k}\frac{1}{2}\sum_{\lambda=1,2}|\vec\epsilon_{\vec k\lambda}\cdot\vec\rho|^2=\frac{1}{3}|\vec\rho|^2
\end{align}
have been used. Note that upon using the energy-conserving $\delta$ function (the last line in Eq.~(\ref{sigma_ab})), the denominators of $A(p,k)$, $B(p,k)$, $C(p,\kappa)$ and $D(p,\kappa)$ defined in Eqs.~(\ref{AB(p,k)}) and (\ref{CD(p,kapp)}) would be $\propto\omega_{\vec\kappa},~\omega_{\vec k}$, respectively, which are simply the corresponding momentum magnitudes if the gluons are massless and thus lead to infrared divergences. In practice, these divergences are regularized and removed by the finite thermal gluon mass in the QGP. For finite gluon mass $m_g$, we define $p_c=\sqrt{(\omega_{\vec k}-\epsilon_B-m_g)m_Q}$. Apparently if $p>p_c$, one has $-\epsilon_B+\omega_{\vec k}-p^2/m_Q<m_g\leq\omega_{\vec\kappa}$, such that the zero point of the argument of the $\delta$ function can never be reached for any $\kappa>0$, and the corresponding integrand in Eq.~(\ref{sigma_ab}) vanishes. Therefore, the $p_c$ acts as an cutoff for the integration over $p$. On the other hand, for $p<p_c$, after integrating out the $\delta$ function via $\int d\kappa$, $\int_0^{p_c}dp$ yields a finite result for the cross section.

For diagrams (c) and (d), similar procedure gives the cross section
\begin{align}\label{sigma_cd}
	&\sigma^{(c+d)}(E_g)=2\pi V \frac{V}{(2\pi)^3}\int d^3p\sum_{b} \frac{V}{(2\pi)^3} \int d^3\kappa\sum_{\sigma}\sum_{c}\nonumber\\  &\times \frac{1}{4\pi}\int d\Omega_{\vec{k}}\frac{1}{2}\sum_{\lambda}\frac{1}{8}\sum_{a} |T_{fi}^{(c)}+T_{fi}^{(d)}|^2\delta(E_i-E_f)\nonumber\\
	&=\frac{g_s^4}{32\pi^4}\frac{1}{E_{g}}\int dp\int \kappa^2\frac{1}{\omega_{\vec{\kappa}}}d\kappa\int d\Omega_{\vec{k}}\int d\Omega_{\vec{\kappa}}\nonumber\\ 	&\Big[\int r^3drj_1(pr)R_{10}(r)\Big]^2 g(\vec{\kappa},\vec{p},\vec{k})\Big[\frac{\omega_{\vec{\kappa}}-\omega_{\vec{k}}}{\omega_{(\vec{\kappa}-\vec{k})}^2-(\omega_{\vec{\kappa}}-\omega_{\vec{k}})^2}\Big]^2 \nonumber\\ 	&\times\delta(-\epsilon_B+\omega_{\vec{k}}-\frac{p^2}{m_Q}-\omega_{\vec{\kappa}}),
\end{align}
where $\sum_{abc}f^{abc}f^{abc}=24$ is used and the polynomial function $g(\vec \kappa,\vec p,\vec k)$ arises from handling the summation over gluon polarizations using the completeness relation Eq.~(\ref{gluon_polarizations_sum}):
\begin{align}\label{g(kappa,p,k)}
	&g(\vec \kappa,\vec p,\vec k)=\sum_{\sigma}\sum_{\lambda}[(\vec p\cdot{\vec \epsilon}_{\vec k\lambda})(\vec k\cdot{\vec \epsilon}_{\vec \kappa\sigma})+(\vec p\cdot{\vec \epsilon}_{\vec \kappa\sigma})(\vec \kappa\cdot{\vec \epsilon}_{\vec k\lambda})\nonumber\\ 	&~~~~~~~~~~~~~~~~~~~~-({\vec \epsilon}_{\vec k\lambda}\cdot{\vec \epsilon}_{\vec \kappa\sigma})X(\vec \kappa,\vec p,\vec k)]^2\nonumber\\
	&=({\vec p}^2-\frac{{(\vec p\cdot\vec k)}^2}{{\vec k}^2})({\vec k}^2-\frac{{(\vec \kappa\cdot\vec k)}^2}{{\vec \kappa}^2})\nonumber\\ 	&+({\vec p}^2-\frac{{(\vec p\cdot\vec \kappa)}^2}{{\vec \kappa}^2})({\vec \kappa}^2-\frac{{(\vec \kappa\cdot\vec k)}^2}{{\vec k}^2})\nonumber\\ 	&+(1+\frac{{(\vec \kappa\cdot \vec k)}^2}{{\vec \kappa}^2{\vec k}^2})X^2(\vec \kappa,\vec p,\vec k)\nonumber\\ 	&+2(\vec p\cdot\vec \kappa-\frac{(\vec p\cdot\vec k)(\vec \kappa\cdot\vec k)}{{\vec k}^2})(\vec p\cdot\vec k-\frac{(\vec p\cdot\vec \kappa)(\vec \kappa\cdot\vec k)}{{\vec \kappa}^2})\nonumber\\ 	&+2\frac{\vec \kappa\cdot\vec k}{{\vec \kappa}^2}(\vec p\cdot\vec \kappa-\frac{(\vec p\cdot\vec k)(\vec \kappa\cdot\vec k)}{{\vec k}^2})X(\vec \kappa,\vec p,\vec k)\nonumber\\ 	&+2\frac{\vec \kappa\cdot\vec k}{{\vec k}^2}(\vec p\cdot\vec k-\frac{(\vec p\cdot\vec \kappa)(\vec k\cdot\vec \kappa)}{{\vec \kappa}^2})X(\vec \kappa,\vec p,\vec k)
\end{align}
with
\begin{align}
	X(\vec \kappa,\vec p,\vec k)=\frac{(\vec p\cdot\vec k)({\vec \kappa}^2-\vec \kappa\cdot\vec k)+(\vec p\cdot\vec \kappa)({\vec k}^2-\vec \kappa\cdot\vec k)}{(\vec \kappa-\vec k)^2}.
\end{align}
When the incident and outgoing gluons are collinear, the denominator in the integrand of Eq.~(\ref{sigma_cd})  $\omega_{(\vec{\kappa}-\vec{k})}^2-(\omega_{\vec{\kappa}}-\omega_{\vec{k}})^2 = -(\omega_{\vec{k}}- \omega_{\vec{\kappa}}-\omega_{(\vec{\kappa}-\vec{k})})(\omega_{\vec{k}}- \omega_{\vec{\kappa}}+\omega_{(\vec{\kappa}-\vec{k})})$ would be vanishing for massless gluons ($|\vec k|-|\vec\kappa|-|\vec k-\vec\kappa|=0$ for $\vec k//\vec\kappa$) and thus a divergence occurs, which will be practically removed by finite gluon masses. In evaluating Eq.~(\ref{sigma_cd}) with Eq.~(\ref{g(kappa,p,k)}), we have chosen the momentum $\vec p$ along the third axis in the momentum space spanned by $\vec k$ and $\vec\kappa$.

Using the completeness relation for spinors Eq.~(\ref{spinor_spin_sum}), the summation and averaging procedure can be performed for the combination of scattering amplitudes from diagrams (e) and (f), to obtain the cross section for $q+J/\psi\to q+c+\bar{c}$
\begin{align}\label{sigma_ef}
	&\sigma^{(e+f)}(E_q) = 2\pi V \frac{V}{(2\pi)^3}\int d^3p\sum_{b} \frac{V}{(2\pi)^3} \int d^3\kappa\sum_{s}\sum_{j}\nonumber\\  &\times \frac{1}{4\pi}\int d\Omega_{\vec{k}}\frac{1}{2}\sum_{r}\frac{1}{3}\sum_{i} |T_{fi}^{(e)}+T_{fi}^{(f)}|^2\delta(E_i-E_f)\nonumber\\
	&=\frac{g_s^4}{72\pi^4}\frac{1}{E_{q}}\int dp\int \kappa^2\frac{1}{E_{\vec{\kappa}}}d\kappa\int d\Omega_{\vec{k}}\int d\Omega_{\vec{\kappa}} \nonumber\\ 	&\Big[\int r^3drj_1(pr)R_{10}(r)\Big]^2 f(\vec{\kappa},\vec{p},\vec{k}) \Big[\frac{E_{\vec{\kappa}}-E_{\vec{k}}}{\omega_{(\vec{\kappa}-\vec{k})}^2-(E_{\vec{\kappa}}-E_{\vec{k}})^2}\Big]^2 \nonumber\\ 	&\times\delta(-\epsilon_B+E_{\vec{k}}-\frac{p^2}{m_Q}-E_{\vec{\kappa}}),
\end{align}
where $E_q=E_{\vec{k}}$ and the polynomial function
\begin{align}
	&f(\vec{\kappa},\vec{p},\vec{k})=m_q^2\sum_{r}\sum_{s}|u^{s\dagger}(\vec{\kappa})\vec{\alpha}\cdot\vec{p}u^{r} (\vec{k})\nonumber\\ 	&-\frac{[u^{s\dagger}(\vec{\kappa})(\vec{\kappa}-\vec{k})\cdot\vec{\alpha}u^{r}(\vec{k})][(\vec{\kappa}-\vec{k})\cdot\vec{p}]}{(\vec{\kappa}-\vec{k})^2}|^2\nonumber\\
	&=2(\vec{p}\cdot\vec{k})(\vec{p}\cdot\vec{\kappa})+\Big[\frac{(\vec{\kappa}-\vec{k})\cdot\vec{p}}{(\vec{\kappa}-\vec{k})^2}\Big]^22(\vec{\kappa}-\vec{k})\cdot\vec{k}(\vec{\kappa}-\vec{k})\cdot\vec{\kappa}\nonumber\\    	&-2\frac{(\vec{\kappa}-\vec{k})\cdot\vec{p}}{(\vec{\kappa}-\vec{k})^2}\Big[(\vec{\kappa}-\vec{k})\cdot\vec{k}(\vec{p}\cdot\vec{\kappa})+(\vec{\kappa}-\vec{k})\cdot\vec{\kappa}(\vec{p}\cdot\vec{k})\Big]\nonumber\\  	&+(\vec{k}\cdot\vec{\kappa}-E_{\vec{k}}E_{\vec{\kappa}}+m_q^2)\Big[-\vec{p}^2+\frac{[(\vec{\kappa}-\vec{k})\cdot\vec{p}]^2}{(\vec{\kappa}-\vec{k})^2}\Big],
\end{align}
When the outgoing quark moves in parallel to the incident quark, $\vec\kappa//\vec k$, and all partons are massless, Eq.~(\ref{sigma_ef}) would have a collinear divergence similar to that identified for Eq.~(\ref{sigma_cd}), which will be practically cured by finite thermal parton masses. The cross section for $\bar{q}+J/\psi\to\bar{q}+c+\bar{c}$ reads the same.

%%%%%%%%%%%%%%%%%%%%%%%%%%%%%%%%%%%%%%%%%%%%%%%%%%%%%%%%%%%%%%%%
\section{Dissociation of various heavy quarkonia in an in-medium potential model}
\label{sec_numerical_diss}
%%%%%%%%%%%%%%%%%%%%%%%%%%%%%%%%%%%%%%%%%%%%%%%%%%%%%%%%%%%%%%%%

%%%%%%%%%%%%%%%%%%%%%%%%%%%%%%%%%%%%%%%%%%%%%%%%%%%%%%%%%%%%%%%%
\subsection{NLO dissociation cross sections}
\label{ssec_numerical_cross_sections}
%%%%%%%%%%%%%%%%%%%%%%%%%%%%%%%%%%%%%%%%%%%%%%%%%%%%%%%%%%%%%%%%

Medium effects enter the dissociation cross sections derived in Sec.~\ref{ssec_diss_cross-sections} through the temperature dependence of bound state wave functions, binding energies and thermal masses of light partons involved. For the in-medium radial wave functions and binding energies of various heavy quarkonium bound states, we have solved the radial Schr$\ddot{o}$digner equation for the $Q\bar{Q}$ system~\cite{Karsch:1987pv,Chen:2017jje}
\begin{align}
&\frac{1}{r^2}\frac{d}{dr}(r^2\frac{dR_{nl}}{dr})+\nonumber \\
&[m_Q(E_{n,l}-2m_Q-V_1(r,T))-\frac{l(l+1)}{r^2}]R_{nl}(r)=0,
\end{align}
with a temperature dependent singlet potential~\cite{Karsch:1987pv}
\begin{equation}\label{V1(r,T)}
V_1(r,T)=-\frac{\alpha}{r}e^{-\mu(T)r}+\frac{\sigma}{\mu(T)}(1-e^{-\mu(T)r}),
\end{equation}
where $\mu/T=-4.058+6.32\cdot(T/T_c-0.885)^{0.1035}$ (critical temperature $T_c$=172.5MeV) parameterizes the screening of the potential~\cite{{Burnier:2015tda}}. The Eq.~(\ref{V1(r,T)}) represents a modification of the vacuum Cornell potential $V_1(r,0)=-\alpha/r+\sigma r$ by color screening, which reproduces well the vacuum masses of various charmonia and bottomonia below threshold with coupling strength $\alpha=4/3\alpha_s=0.471$, string tension $\sigma=0.192$\,GeV$^2$ and charm and bottom quark masses $m_c=1.320$\,GeV and $m_b=4.746$\,GeV, respectively~\cite{Karsch:1987pv}. The temperature-dependent binding energy of a bound state is then obtained via $\epsilon_B(T)=2m_Q+\sigma/\mu(T)-E_{n,l}(T)$, whose zero point defines the melting temperature for the bound state~\cite{Karsch:1987pv,Chen:2017jje}. For example, the melting temperatures $T_d\simeq1.6T_c,~1.2T_c$ for $J/\psi$, $\chi_c$ (and $\psi(2S)$), respectively, and $T_d\simeq2.6T_c,~1.5T_c,~1.3T_c$ for $\Upsilon(1S)$, $\Upsilon(2S)$ (and $\chi_b(1P)$) and $\Upsilon(3S)$, respectively.

For the thermal masses of gluons and light quarks/antiquarks that help to remove the infrared and collinear divergences as discussed in Sec.~\ref{ssec_diss_cross-sections}, we use their perturbative values $m_g(T)=\sqrt{3/4}g_sT$ and $m_q(T)=\sqrt{1/3}g_sT$ with fixed $g_s=2.3$ for three active light flavors~\cite{Cao:2018ews}. However, it was noted~\cite{Gossiaux:2008jv,Zhao:2023ucp} that to mimic the same partonic energy loss from some nonperturbative calculations~\cite{Braaten:1991jj} with a Born diagram, the screening on the exchanged gluon propagator in the $t$-channel scattering should be much reduced. In this spirit, we use a significantly reduced effective mass $m_{g_{\rm ex}}(T)=r g_sT$, with $r=0.2$~\cite{Zhao:2023ucp}, for the intermediate gluon associated with momentum $\vec k-\vec\kappa$ in diagrams (c), (d), (e) and (f), that would correspond to the exchanged gluon in the $t$-channel Born diagram in relativistically covariant perturbation theory, such that $\omega_{(\vec k-\vec\kappa)}=\sqrt{(\vec k-\vec\kappa)^2+m_{g_{\rm ex}}^2}$ in Eqs.~({\ref{sigma_cd}}) and (\ref{sigma_ef}). In this way, we expect to partially account for the pertinent nonperturabtive effects in the current perturbative calculations.

\begin{figure}[!t]
    \includegraphics[width=1.0\columnwidth]{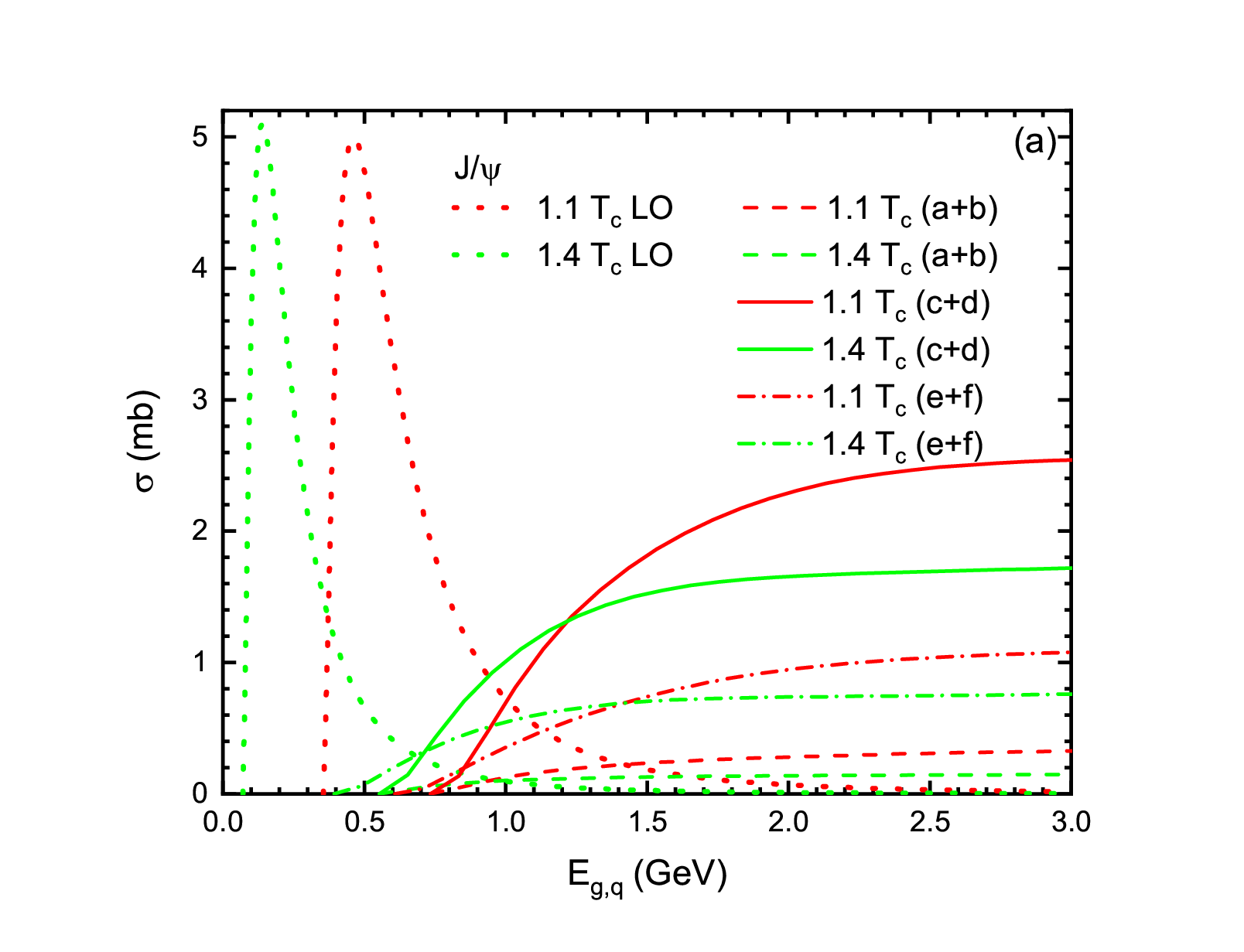}
	\vspace{-0.05cm}
    \includegraphics[width=1.0\columnwidth]{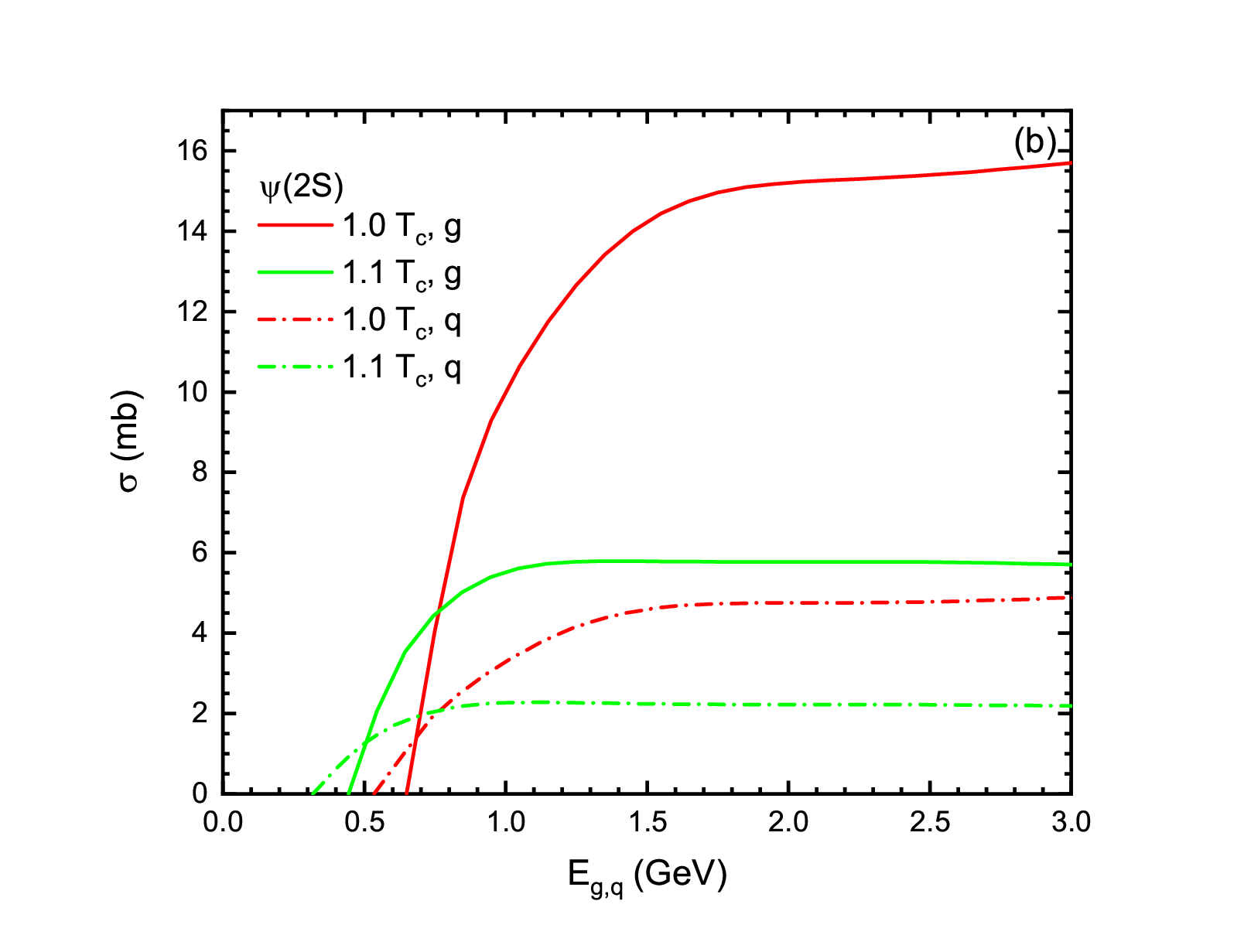}
	\vspace{-0.05cm}
    \includegraphics[width=1.0\columnwidth]{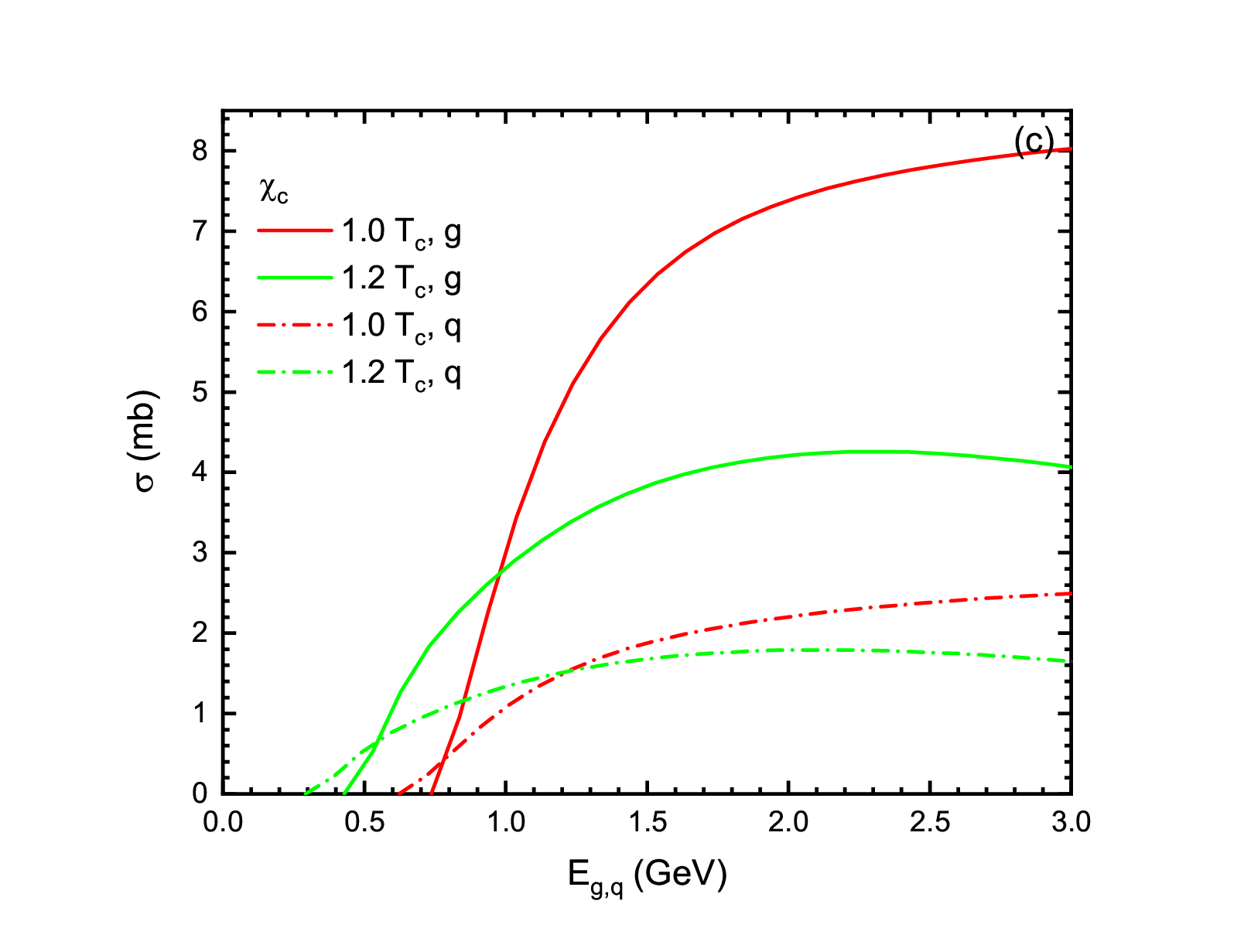}
	\vspace{-0.05cm}
	\caption{NLO dissociation cross sections for charmonia at varying temperatures. For (a) $J/\psi$, $\sigma^{(a+b)}$, $\sigma^{(c+d)}$ and $\sigma^{(e+f)}$ are displayed separately, and LO (gluo-dissociation) cross sections taken from~\cite{Chen:2017jje} are also shown for comparison. For (b) $\psi(2S)$ and (c) $\chi_c$, NLO dissociation cross sections by gluons ($\sigma^{(a+b)}+\sigma^{(c+d)}$) and quarks ($\sigma^{(e+f)}$) are separately shown.}
	\label{figs_charmonia_diss_cross_section}
\end{figure}

\begin{figure}[!t]
    \includegraphics[width=1.0\columnwidth]{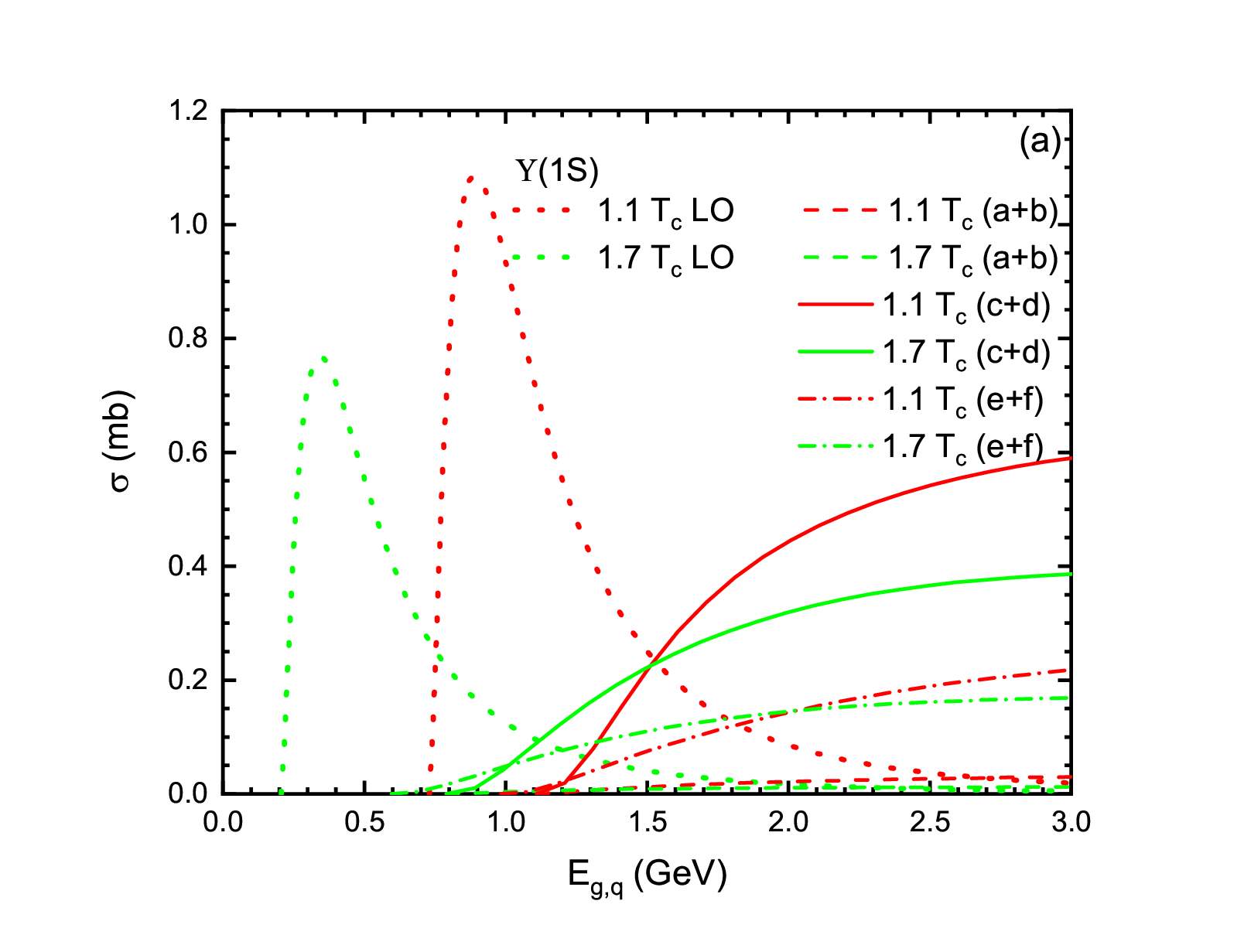}
	\vspace{-0.05cm}
    \includegraphics[width=1.0\columnwidth]{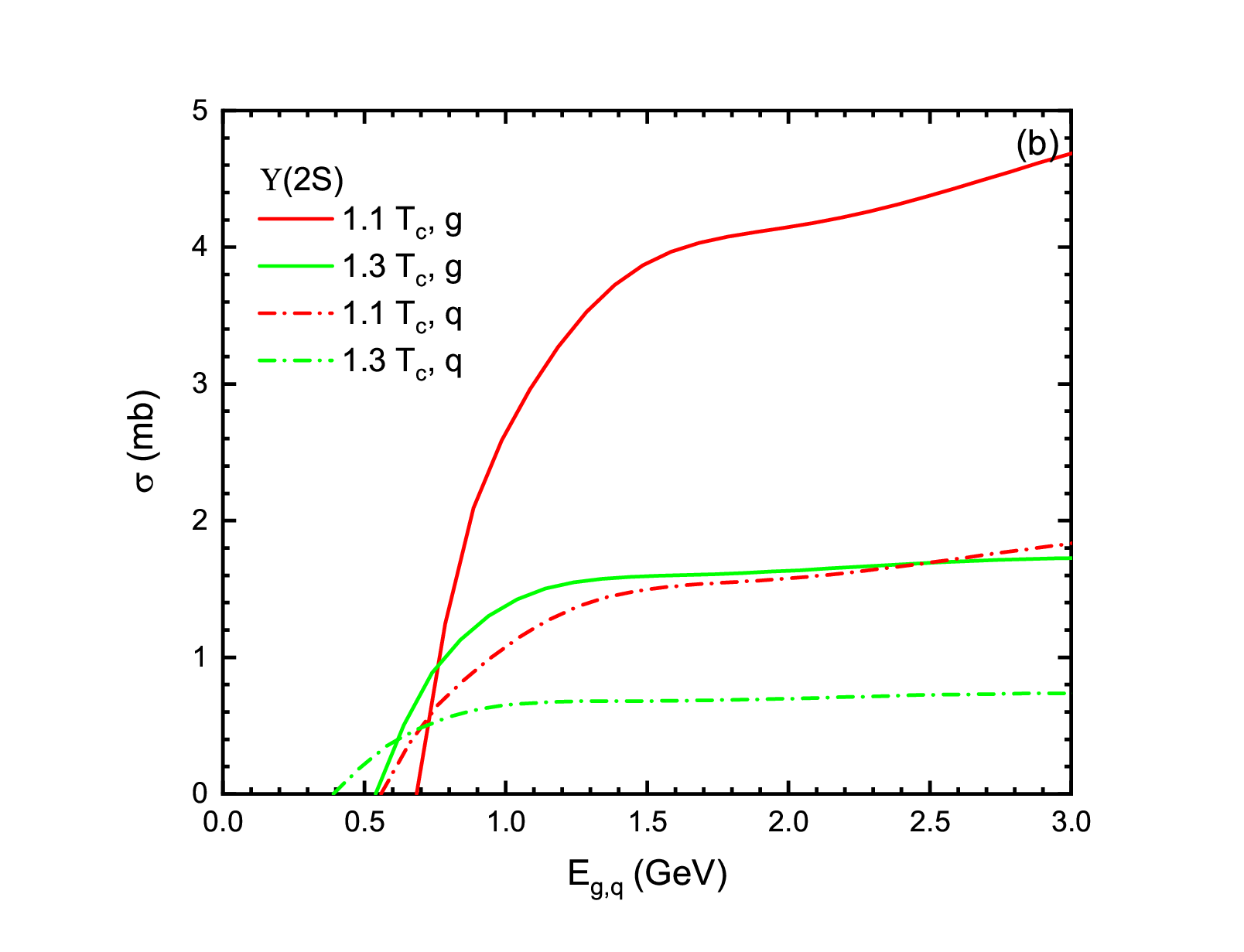}
	\vspace{-0.05cm}
    \includegraphics[width=1.0\columnwidth]{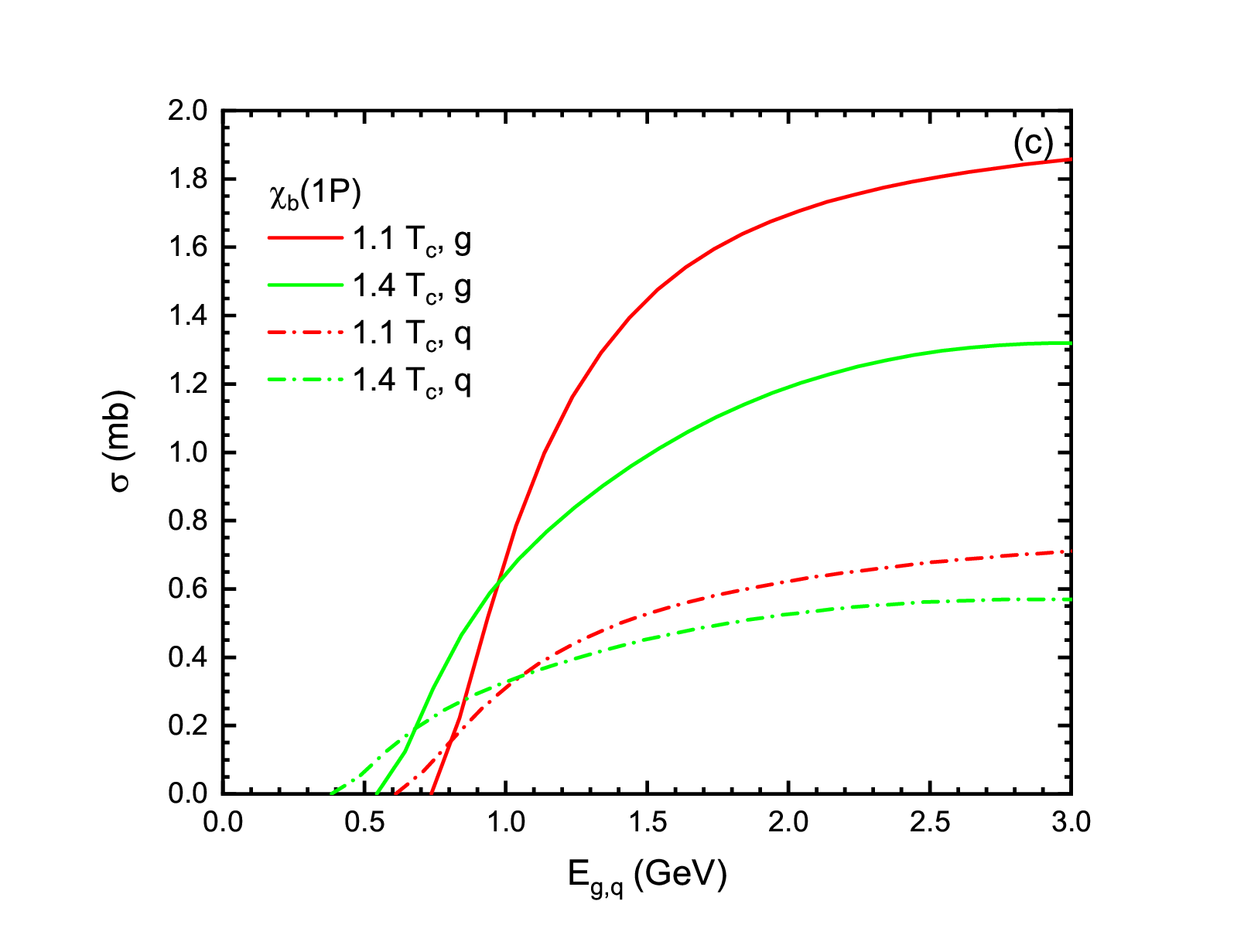}
	\vspace{-0.05cm}
	\caption{Same as Fig.~\ref{figs_charmonia_diss_cross_section} but for bottomonia: (a) $\Upsilon(1S)$, (b) $\Upsilon(2S)$ and (c) $\chi_b(1P)$.} %For (a) $\Upsilon(1S)$, $\sigma^{(a+b)}$, $\sigma^{(c+d)}$ and $\sigma^{(e+f)}$ are displayed separately, and LO (gluo-dissociation) cross sections taken from~\cite{Chen:2017jje} are also shown for comparison. For (b) $\Upsilon(2S)$ and (c) $\chi_b$, NLO dissociation cross sections by gluons ($\sigma^{(a+b)}+\sigma^{(c+d)}$) and quarks ($\sigma^{(e+f)}$) are shown.}
	\label{figs_bottomonia_diss_cross_section}
\end{figure}

The numerical results of NLO dissociation cross sections for various charmonia and bottomonia are displayed in Fig.~\ref{figs_charmonia_diss_cross_section} and Fig.~\ref{figs_bottomonia_diss_cross_section}, respectively, as a function of the incident parton energy at varying temperatures. A general observation is that the NLO cross sections first grow with the incident parton energy and then saturate toward large energies, in marked contrast to the LO (gluo-dissocation) cross sections which exhibit a pronounced peak at energy (wavelength) matching the bound state binding energy (size) and quickly drop off thereafter, as shown in Fig.~\ref{figs_charmonia_diss_cross_section}(a) and Fig.~\ref{figs_bottomonia_diss_cross_section}(a) for the $1S$ states $J/\psi$ and $\Upsilon(1S)$, respectively. This is presumably attributed to the outgoing parton in the final state of the NLO processes carrying away the excess energy, thereby overcoming the mismatch between the incident parton wavelength and the bound state size in the LO process. Among the three types of NLO dissociation cross sections as derived in Sec.~\ref{ssec_diss_cross-sections}, $\sigma^{(a+b)}$ turns out much smaller than $\sigma^{(c+d)}$ and $\sigma^{(e+f)}$, as demonstrated for $J/\psi$ and $\Upsilon(1S)$. This may be expected from the fact that what is exchanged in $s$- and $u$-channel-like diagrams (a) and (b) (Fig.~\ref{figs[a]+[b]}) is a very massive $(Q\bar{Q})_8$ octet, whereas underlying $\sigma^{(c+d)}$ and $\sigma^{(e+f)}$ are the $t$-channel-like one-gluon-exchange diagrams (c) and (d) (Fig.~\ref{figs[c]+[d]}), and (e) and (f) (Fig.~\ref{figs[e]+[f]}). As temperature increases, the bound state wave functions (dipole size) expand and the binding energies decrease. While the latter is responsible for the shift of the onset point of cross sections toward lower incident parton energy, the former is expected to enhance the magnitude of cross sections. However, this naive expectation is quantitatively offset by the suppression from the increasing effective mass of the exchanged gluon $m_{g_{\rm ex}}$ (as well as the decreasing binding energy parameter entering the cross section formulas, as we've numerically checked), leading to an overall reduction of cross sections toward higher temperatures.

Comparing different heavy quarkonium states, the NLO cross sections are roughly ordered by their vacuum binding energies: more loosely bound excited states (a dipole of larger size) are more easily dissociated by thermal partons and thus possess larger dissociation cross sections. The NLO dissociation cross sections for Bottomonia are consistently smaller than those for charmonia. The $\Upsilon(1S)$ is the most tightly bound state, for which the color dipole coupling mechanism underlying our calculations may be most applicable. Furthermore, the technical approximation made in our calculations that the rest frame of the heavy quarkonium is also considered to be the rest frame of the unbound octet $(Q\bar{Q})_8$ in the final state, {\it i.e.}, the recoil effect on the bound state by the incident parton is neglected, should be more reasonable for the much more massive bottomonia. Therefore, our results of the NLO dissociation cross sections may be deemed most reliable for the $\Upsilon(1S)$. To compare with the corresponding result shown in~\cite{Hong:2018vgp}, where the NLO cross sections for $\Upsilon(1S)$ were calculated from another perturbative approach using an effective vertex constructed from the non-relativistic Bethe-Salpeter amplitude, we choose a comparison point of $T\sim300$\,MeV and incident parton energy $\sim 1.5$\,GeV, at which our value of $\sigma(g+\Upsilon(1S)\to g+b+\bar{b})\sim 0.2$\,mb is a factor of $\sim 2$ larger, which might be in part due to the use of a reduced effective mass ($m_{g_{\rm ex}}$) for the exchanged gluon in our calculations. However, the cross section for $\Upsilon(1S)$ shown in~\cite{Hong:2018vgp} keeps increasing rapidly with the incident parton energy, differing remarkably from the near-saturation of the NLO cross section toward higher energies as found in our calculations.

%%%%%%%%%%%%%%%%%%%%%%%%%%%%%%%%%%%%%%%%%%%%%%%%%%%%%%%%%%%%%%%%
\subsection{Dissociation rates}
\label{ssec_diss_rates}
%%%%%%%%%%%%%%%%%%%%%%%%%%%%%%%%%%%%%%%%%%%%%%%%%%%%%%%%%%%%%%%%

The dissociation rates, serving as inputs for the phenomenological modelling of heavy quarknonium transport in the QGP~\cite{Zhao:2010nk,Song:2011nu,Strickland:2011mw,Zhou:2014kka,Du:2017qkv,Brambilla:2020qwo,Wu:2024gil}, are obtained by folding the dissociation cross section with the distribution function of the incident parton. For a bound state sitting at rest in the QGP, the dissociation rate reads
\begin{align}\label{dissocation_rates}
	\Gamma(T)=d_{g,q}\int \frac{d^3k}{(2\pi)^3}f_{g,q}(E_{g,q}(\vec k),T)v_{\rm rel}\sigma(E_{g,q},T),
\end{align}
where $d_g=2\cdot8=16$ for thermal gluons and $d_q=3\cdot2\cdot3\cdot2=36$ for thermal ligth quarks/antiquarks is the incident parton degeneracy, $v_{\rm rel}$ the relative velocity between the incident parton and the heavy quarkonium at rest, and $f_{g,q}(E_{g,q},T)=1/(e^{(E_{g,q}(\vec{k})/T)}\mp 1)$ the Bose ($-$)/Fermi ($+$) distribution with $E_{g,q}=\sqrt{\vec k^2+m_{g,q}^2(T)}$ for gluons and light quarks/antiquarks, respectively. We note that for typical temperatures (a couple of hundreds of MeV) reached in the QGP created in current relativistic heavy-ion collisions, the thermal parton energy is on the order of $\leq 1$\,GeV, which is much smaller than the mass of charmonia and bottomonia. Therefore, in calculating the dissociation rate, thermal partons are generally not probing the NLO cross sections at very high energies, which also renders the approximation of neglecting the recoil effect on the heavy quarkonium relatively safe.

\begin{figure}[!t]
    \includegraphics[width=1.0\columnwidth]{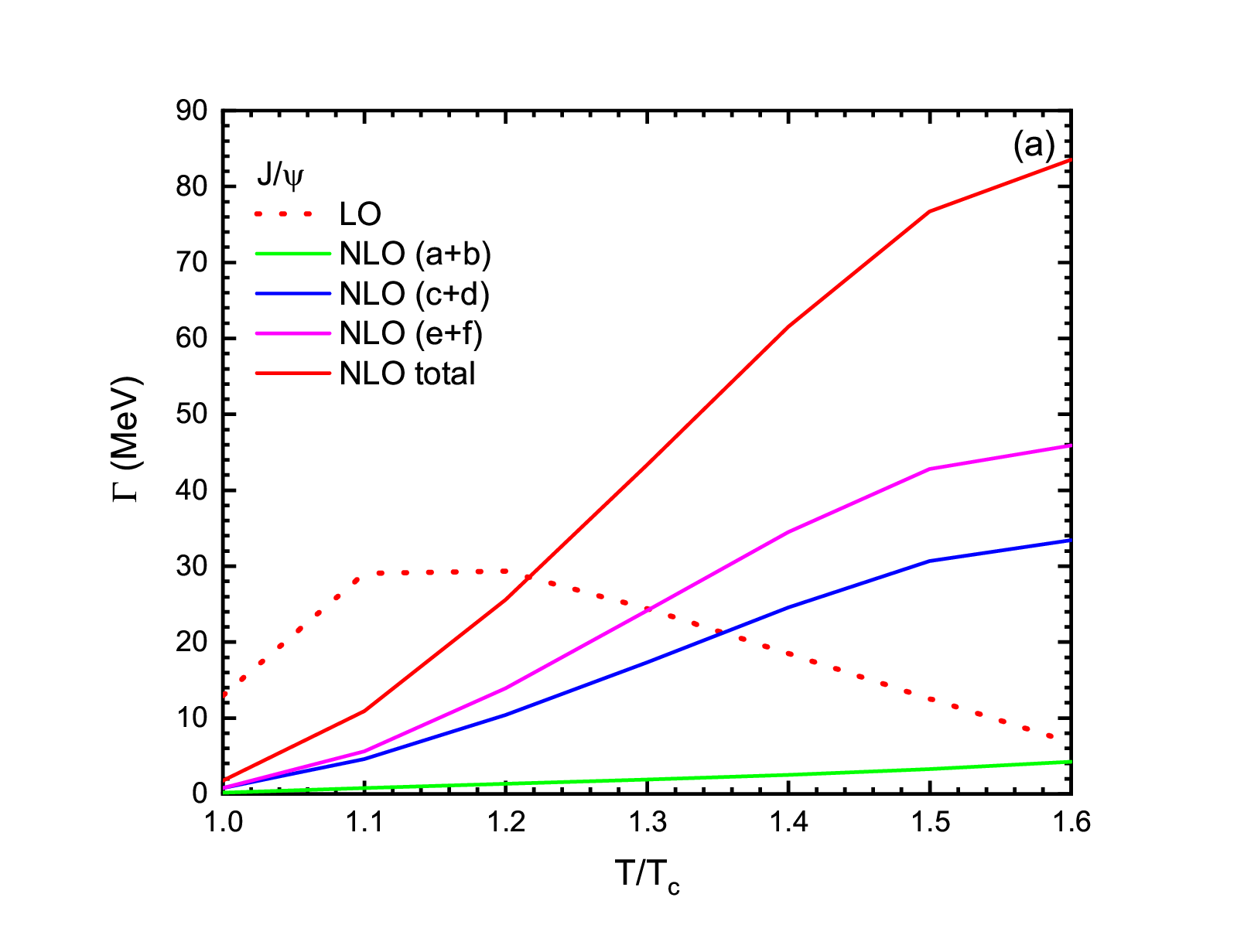}
	\vspace{-0.05cm}
    \includegraphics[width=1.0\columnwidth]{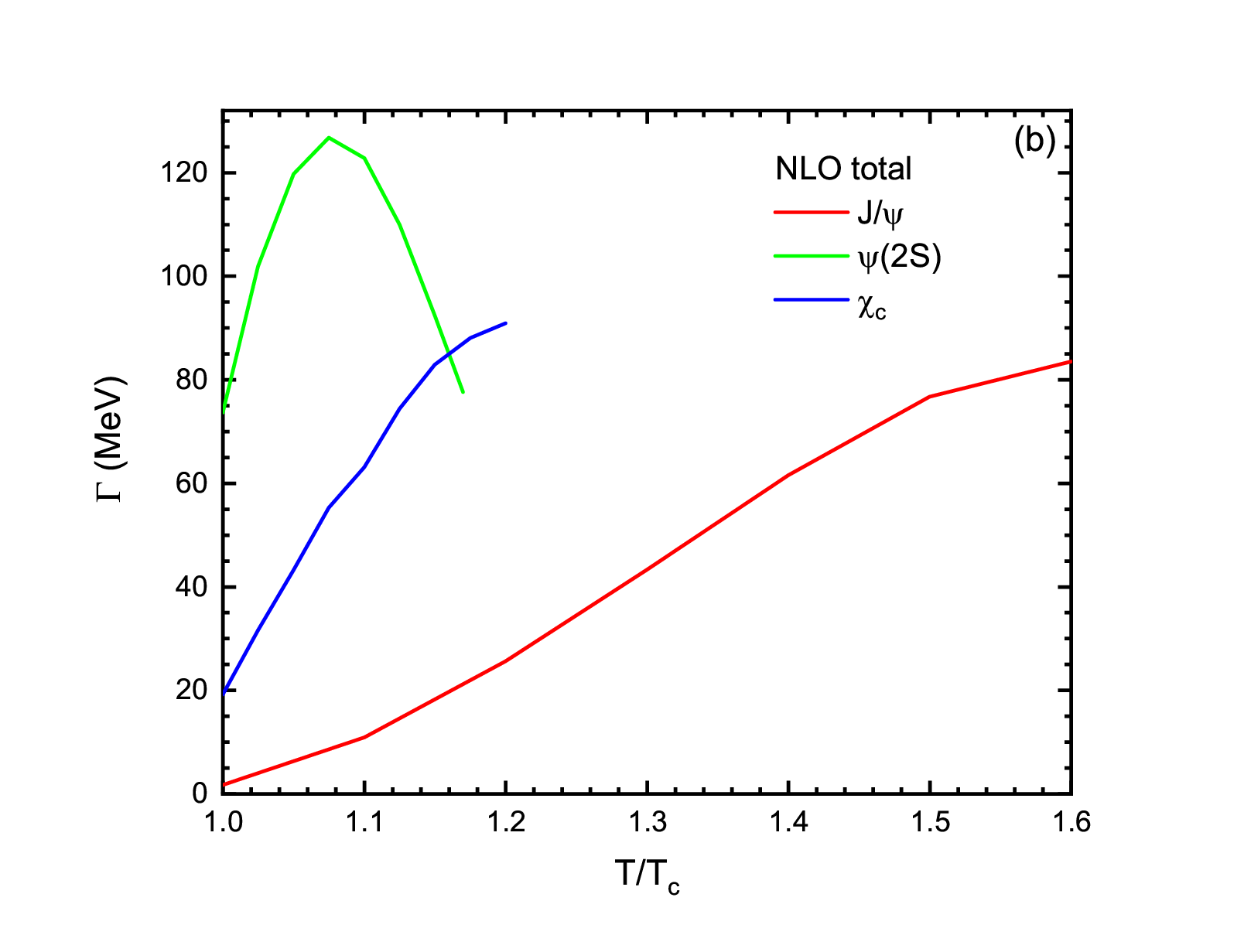}
	\vspace{-0.05cm}
	\caption{(a) NLO dissociation rates for $J/\psi$ at finite temperatures; contributions from different NLO processes and from LO (gluo-dissociation)~\cite{Chen:2017jje} process are separately displayed for comparison. (b) Total NLO dissociation rates for various charmonia.}
	\label{figs_diss_rate_Charmonia}
\end{figure}

\begin{figure}[!t]
    \includegraphics[width=1.0\columnwidth]{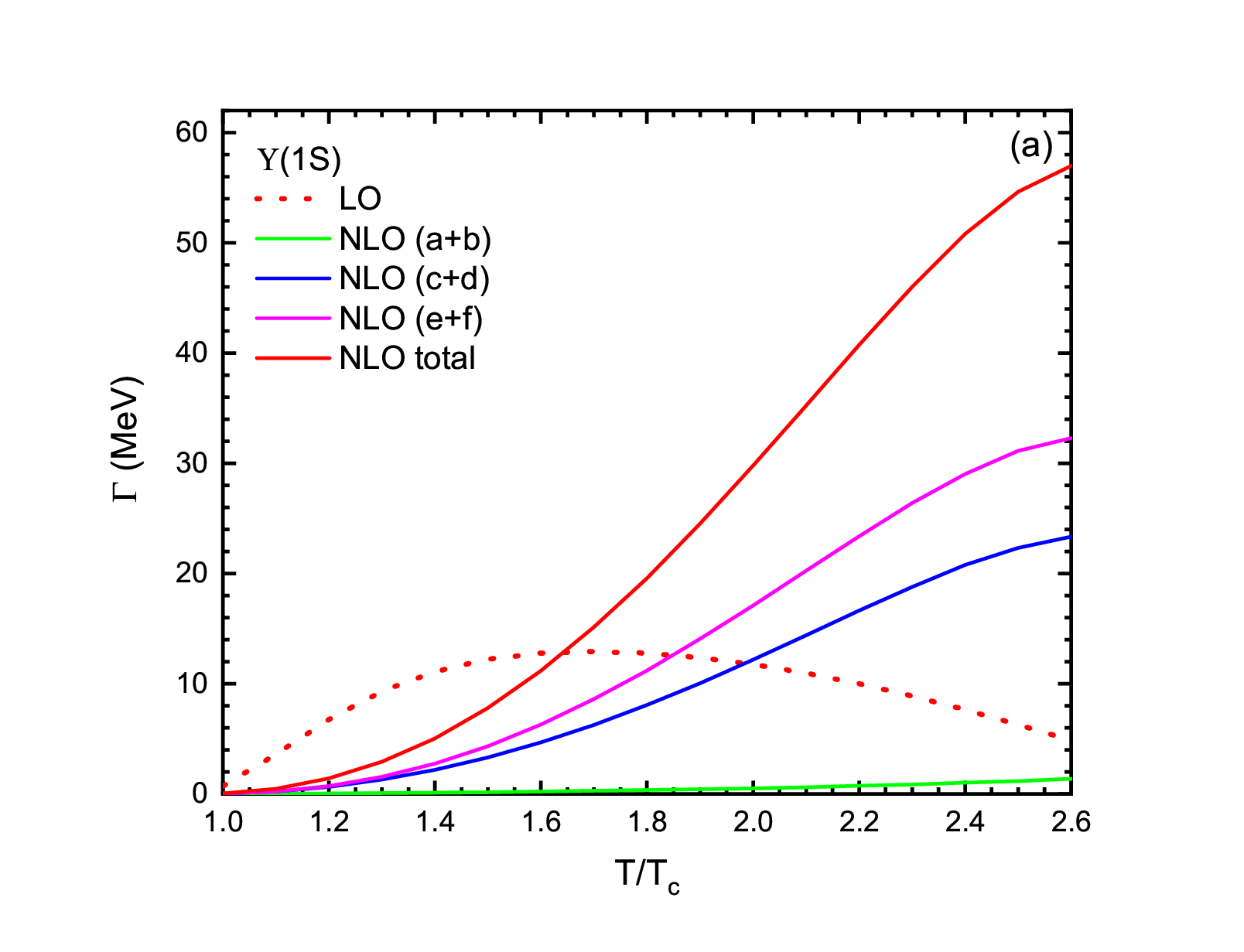}
	\vspace{-0.05cm}
    \includegraphics[width=1.0\columnwidth]{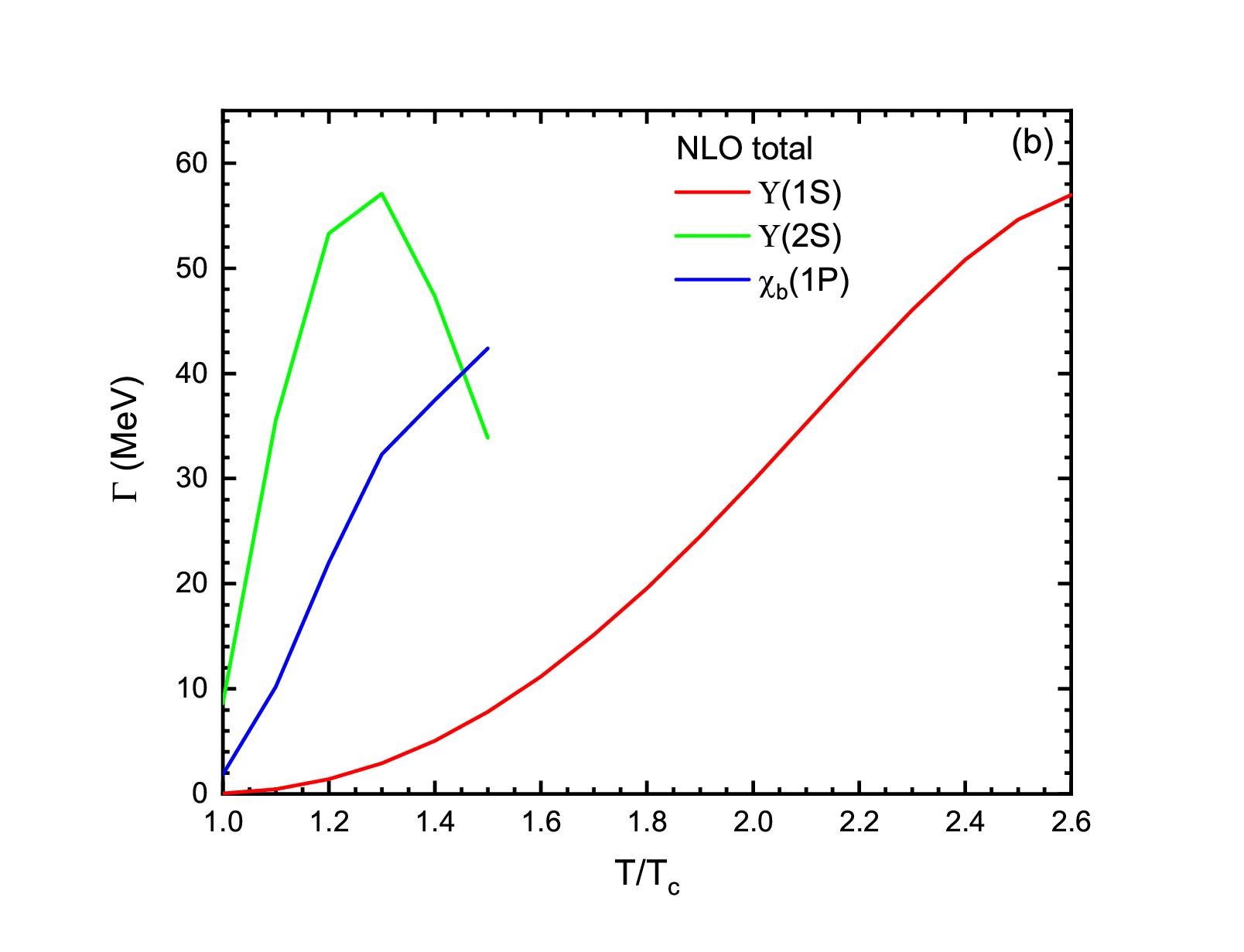}
	\vspace{-0.05cm}
	\caption{(a) NLO dissociation rates for $\Upsilon(1S)$ at finite temperatures; contributions from different NLO processes and from LO (gluo-dissociation) process are separately displayed for comparison. (b) Total NLO dissociation rates for various bottomonia.}
	\label{figs_diss_rate_Bottomonia}
\end{figure}

The calculated dissociation rates for charmonia and bottomonia are presented in Figs.~\ref{figs_diss_rate_Charmonia} and~\ref{figs_diss_rate_Bottomonia} up to their respective melting temperatures. The dissociation rates from different NLO processes are displayed separately and compared to the counterpart from LO (gluo-dissociation) process (taken from~\cite{Chen:2017jje}) for the $1S$ groud states $J/\psi$ (Fig.~\ref{figs_diss_rate_Charmonia}(a)) and $\Upsilon(1S)$ (Fig.~\ref{figs_diss_rate_Bottomonia}(a)). At low temperatures (up to $\sim 1.2T_c$ and $\sim1.6T_c$ for $J/\psi$ and $\Upsilon(1S)$, respectively), the ground states are still sufficiently tightly bound and the incident parton of long enough wavelength does not resolve the substructure of the bound state, so that the LO dissociation remains more effective and its dissociation rates dominate over the total NLO results. As temperature increases, the peak of the LO dissociation cross section shifts toward lower energies (cf. Fig.~\ref{figs_charmonia_diss_cross_section}(a) and Fig.~\ref{figs_bottomonia_diss_cross_section}(a)), where the incident parton has a smaller and smaller phase space~\cite{Rapp:2008tf}, leading to decreasing dissociation rates. In contrast, the NLO dissociation rates keep growing with temperature and take over from the LO counterparts toward high temperatures. The total NLO dissociation rates for $\Upsilon(1S)$ from our calculation turn out to be quite comparable to corresponding results in~\cite{Hong:2018vgp} as computed from another approach. However, we note that in~\cite{Hong:2018vgp}, the dissociation rates were calculated using massless incident parton (therefore of significantly higher density) with a smaller NLO cross section, whereas we have taken into account the thermal masses of the partons colliding with the heavy quarkonium.

The total NLO dissociation rates for the excited charmonia and bottomonia as shown in Fig.~\ref{figs_diss_rate_Charmonia}(b) and Fig.~\ref{figs_diss_rate_Bottomonia}(b), respectively, are generally larger than their ground state counterparts at the same temperatures, in accord with the larger dissociation cross sections for these states. Here, we note that for $\psi(2S)$ and $\Upsilon(2S)$, the NLO dissociation rates turn out to drop off when approaching their respective melting temperatures, in contrast to the monotonously increasing behavior for the $1S$ and $1P$ states. This is due to the fast decline of their NLO cross sections with temperature (cf. Fig.~\ref{figs_charmonia_diss_cross_section}(b) and Fig.~\ref{figs_bottomonia_diss_cross_section}(b)), which, in turn, can be technically attributed to the peculiar behavior of their radial functions having a node, as we've numerically checked.

%%%%%%%%%%%%%%%%%%%%%%%%%%%%%%%%%%%%%%%%%%%%%%%%%%%%%%%%%%%%%%%%
\section{Second-order transition between different eigen-bound states}
\label{sec_transition}
%%%%%%%%%%%%%%%%%%%%%%%%%%%%%%%%%%%%%%%%%%%%%%%%%%%%%%%%%%%%%%%%

%%%%%%%%%%%%%%%%%%%%%%%%%%%%%%%%%%%%%%%%%%%%%%%%%%%%%%%%%%%%%%%%
\subsection{Deriving the transition cross sections}
\label{ssec_transtion_cross-section}
%%%%%%%%%%%%%%%%%%%%%%%%%%%%%%%%%%%%%%%%%%%%%%%%%%%%%%%%%%%%%%%%

In addition to the NLO dissociation, the heavy quarkonium may experience quantum transition to another eigen-bound state in the QGP (including excitation and elastic scattering), which also occurs at the second order as required by color neutrality, by first absorbing and then emitting a gluon. This process may also contribute to the total thermal decay rates for heavy quarknoium as evaluated in lattice QCD~\cite{Larsen:2019zqv}. We compute the cross sections and rates for these second-order transition processes in the same framework as for the NLO dissociation. Using the vertex $V_{SO}$ twice, the Feynamn diagrams for the transition $g+\Psi \to g+\Psi'$ ($\Psi$ and $\Psi'$ denote two eigen-bound states of heavy quarkonium) can be constructed, as illustrated in Fig.~\ref{figs_transition} for $g+J/\psi\to g+\psi(2S)$, which looks rather similar to the dissociation diagrams in Fig.~\ref{figs[a]+[b]}, except that now the final state involves another bound state instead of the unbound octet. We note that a $s$-wave bound state can only transition into another $s$-wave or $d$-wave bound state, {\it i.e.}, the change of orbital angular momentum is constrained to be even, as dictated by the selection rules ($\Delta l=1, \Delta s=0$) for the color-electric dipole coupling~\cite{Chen:2017jje}.

\begin{figure} [!t]
\includegraphics[width=1.0\columnwidth]{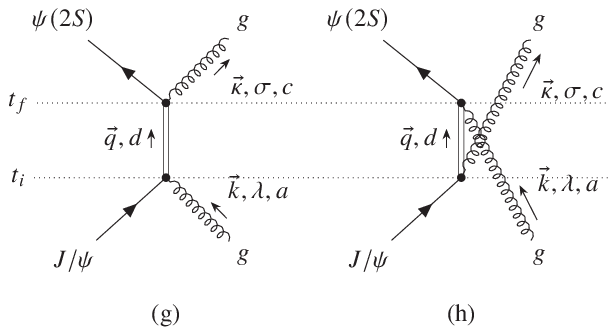}
\vspace{-0.3cm}
\caption{Feynman diagrams for $g+J/\psi\to g+\psi(2S)$. The time direction goes upward. Between the initial time $t_i$ and final time $t_f$ is the intermediate state.}
\label{figs_transition}
\end{figure}

For the transition represented in Fig.~\ref{figs_transition}, the initial state $|i\rangle=| J/\psi,  g(\vec{k},\lambda,a) \rangle$ and final state $|f\rangle=| \psi(2S),  g(\vec{\kappa},\sigma,c) \rangle$. For diagram (g), the intermediate state reads $| m \rangle=|(c\bar c)_8(\vec q,d)\rangle$. The transition matrix element $\langle m |V_{SO}| i \rangle$ is the same as the Eq.~(\ref{V_SO^a}), and the other one
\begin{align}
	& \langle f |V_{SO}| m \rangle = \langle \psi(2S),  g(\vec{\kappa},\sigma,c) |V_{SO}| (c\bar c)_8(\vec q,d) \rangle \nonumber\\
  	&=g_s\frac{\delta^{dc}}{V}\sqrt{\frac{\pi\omega_{\vec{\kappa}}}{3}}e^{-i\vec{\kappa}\cdot\vec{x}} (\vec{\epsilon}_{\vec{\kappa}\sigma}\cdot\vec{q})\frac{1}{q}\int r^3drj_1(qr)R_{20}(r).
\end{align}
By carrying out the summation over intermediate state $\sum_{m}=\sum_{\vec{q}}\sum_{d} = \frac{V}{(2\pi)^3}\int d^{3}q \sum_{d}$, the transition amplitude is obtained
\begin{align}
	T_{fi}^{(g)}&=\sum_{m}\frac{\langle f |V_{SO}| m \rangle \langle m |V_{SO}| i  \rangle}{E_i-E_m+i\epsilon}\nonumber\\	&=\delta^{ac}\frac{g_s^2\pi}{3V}\sqrt{\omega_{\vec{\kappa}}\omega_{\vec{k}}}e^{i(\vec{k}-\vec{\kappa})\cdot\vec{x}}\int\frac{d^3q}{(2\pi)^3}\Big[\frac{(\vec{\epsilon}_{\vec{k}\lambda}\cdot\vec{q})(\epsilon_{\vec{\kappa}\sigma}\cdot\vec{q})}{q^2}\nonumber\\ 	&\times\frac{\int r^3drj_1(qr)R_{20}(r)\times\int r^3drj_1(qr)R_{10}(r)}{-\epsilon_B^{J/\psi}+\omega_{\vec{k}}-\frac{q^2}{m_Q}+i\epsilon}\Big].
\end{align}

For diagram (h), the intermediate state is $ | m \rangle=| (c\bar c)_8(\vec q,d),\,g(\vec{k},\lambda,a),\, g(\vec{\kappa},\sigma,c) \rangle$. The transition matrix element $\langle m |V_{SO}| i \rangle$ is the same as Eq.~(\ref{V_SO^b}), and the other one reads
\begin{align}
	& \langle f |V_{SO}| m \rangle = \langle \psi' |V_{SO}| (c\bar c)_8(\vec q,d),\,g(\vec{k},\lambda,a) \rangle \nonumber\\
	&=(-)g_s\frac{\delta^{da}}{V}\sqrt{\frac{\pi\omega_{\vec{k}}}{3}}e^{i\vec{k}\cdot\vec{x}} (\vec{\epsilon}_{\vec{k}\lambda}\cdot\vec{q})\frac{1}{q}\int r^3drj_1(qr)R_{20}(r).
\end{align}
Upon summing over the intermediate states, the transition amplitude reads
\begin{align} T_{fi}^{(h)}&=\delta^{ac}\frac{g_s^2\pi}{3V}\sqrt{\omega_{\vec{\kappa}}\omega_{\vec{k}}}e^{i(\vec{k}-\vec{\kappa})\cdot\vec{x}}\int\frac{d^3q}{(2\pi)^3}\Big[\frac{(\vec{\epsilon}_{\vec{k}\lambda}\cdot\vec{q})(\vec{\epsilon}_{\vec{\kappa}\sigma}\cdot\vec{q})}{q^2}\nonumber\\ 	&\times\frac{\int r^3drj_1(qr)R_{20}(r)\times\int r^3drj_1(qr)R_{10}(r)}  {-\epsilon_B^{J/\psi}-\frac{q^2}{m_Q}-\omega_{\vec{\kappa}}+i\epsilon}\Big].
\end{align}

Combining these two transition amplitudes, the cross section for $g+J/\psi\to g+\psi(2S)$ is obtained via the standard averaging and summation procedure
\begin{align}\label{sigma_gpsi-to-gpsi2S}
	&\sigma^{(g+h)}(E_g)= 2\pi V \frac{V}{(2\pi)^3} \int d^3\kappa\sum_{\sigma}\sum_{c} \frac{1}{4\pi} \int d\Omega_{\vec{k}}\frac{1}{2}\sum_{\lambda}\frac{1}{8}\sum_{a} \nonumber\\  & ~~~~~~~~~~~~ \times|T_{fi}^{(g)}+T_{fi}^{(h)}|^2\delta(E_i-E_f)\nonumber\\
	&=\frac{{g_s}^4}{2^{5}3^{4}{\pi}^5}\omega_{\vec k} \int_0^{\infty}d\kappa {\kappa}^2{\omega}_{\vec\kappa} Tr[\mathcal{M}^2] \delta (-{\epsilon}_B^{J/\psi}+{\omega}_{\vec k}+{\epsilon}_B^{\psi(2S)}-{\omega}_{\vec \kappa}),
\end{align}
with the matrix $\mathcal{M}$ being
\begin{align}
	& \mathcal{M}_{ij}=\int {d^3 q}\Big[ \frac{q_i q_j}{q^2} {\int r^3 dr j_1(qr) R_{10}(r) \int  r^3 dr j_1(qr) R_{20}(r)}\nonumber\\
	&\times ( \frac{1}{-\epsilon_B^{J/\psi}+\omega_{\vec k}-\frac{q^2}{m_Q}+i\epsilon} + \frac{1}{-\epsilon_B^{J/\psi}-\frac{q^2}{m_Q} -\omega_{\vec \kappa}+i\epsilon})\Big],
\end{align}
where $i, j$=1, 2, 3 label the Cartesian momentum components. It is not hard to prove that all off-diagonal elements vanish and one is left with only the diagonal elements
\begin{align}\label{M_ii}
	&\mathcal{M}_{11}=\mathcal{M}_{22}=\mathcal{M}_{33}\nonumber\\
	&=\frac{4\pi}{3}\int_0^{\infty} dq \Big[ q^2 {\int r^3 dr j_1(qr) R_{10}(r) \int  r^3 dr j_1(qr) R_{20}(r)}\nonumber\\
	&\times( \frac{1}{-\epsilon_B^{J/\psi}+\omega_{\vec k}-\frac{q^2}{m_Q}+i\epsilon}+\frac{1}{-\epsilon_B^{J/\psi} -\frac{q^2}{m_Q}-\omega_{\vec \kappa} +i\epsilon}) \Big],
\end{align}
where, $q=|\vec q|$ is the magnitude of the internal relative momentum of intermediate octet $(c\bar{c})_8$. We use the analytical Coulomb radial wave function to mimic the numerically solved radial function with full potential (Eq.~(\ref{V1(r,T)})) by tuning the Bohr radius parameter. This way, the two integrations over radial wave functions in Eq.~(\ref{M_ii}) can be analytically performed and what's left is two polynomial functions of $q$
\begin{align}\label{q-polynomial}
\frac{2^4a^{7/2}q}{(1+a^2q^2)^3},~~~~~~\frac{2^{19/2}a^{7/2}q(2a^2q^2-1)}{(1+4a^2q^2)^4},
\end{align}
with $a$ being the Bohr radius parameter. These two polynomials contain higher order poles with respect to $q$, which, together with the pole $q= \sqrt {m_{Q} ( \omega_{\vec{k}} - \epsilon_{B}^{J/\psi})}$ in the first term of the third line of Eq.~(\ref{M_ii}) when $\omega_{\vec{k}} > \epsilon_{B}^{J/\psi}$, can be handled by extending the integration to the upper half plane of complex $q$ (noting that the integrand is an even function with respect to $q$) and then making use of the residual theorem.

We've also derived the transition cross sections between two $p$-wave bound states. These cross sections, including Eq.~(\ref{sigma_gpsi-to-gpsi2S}), suffer from no divergence.

%%%%%%%%%%%%%%%%%%%%%%%%%%%%%%%%%%%%%%%%%%%%%%%%%%%%%%%%%%%%%%%%
\subsection{Numerical results}
\label{ssec_transtion_numerical-results}
%%%%%%%%%%%%%%%%%%%%%%%%%%%%%%%%%%%%%%%%%%%%%%%%%%%%%%%%%%%%%%%%

\begin{figure}[!t]
    \includegraphics[width=1.0\columnwidth]{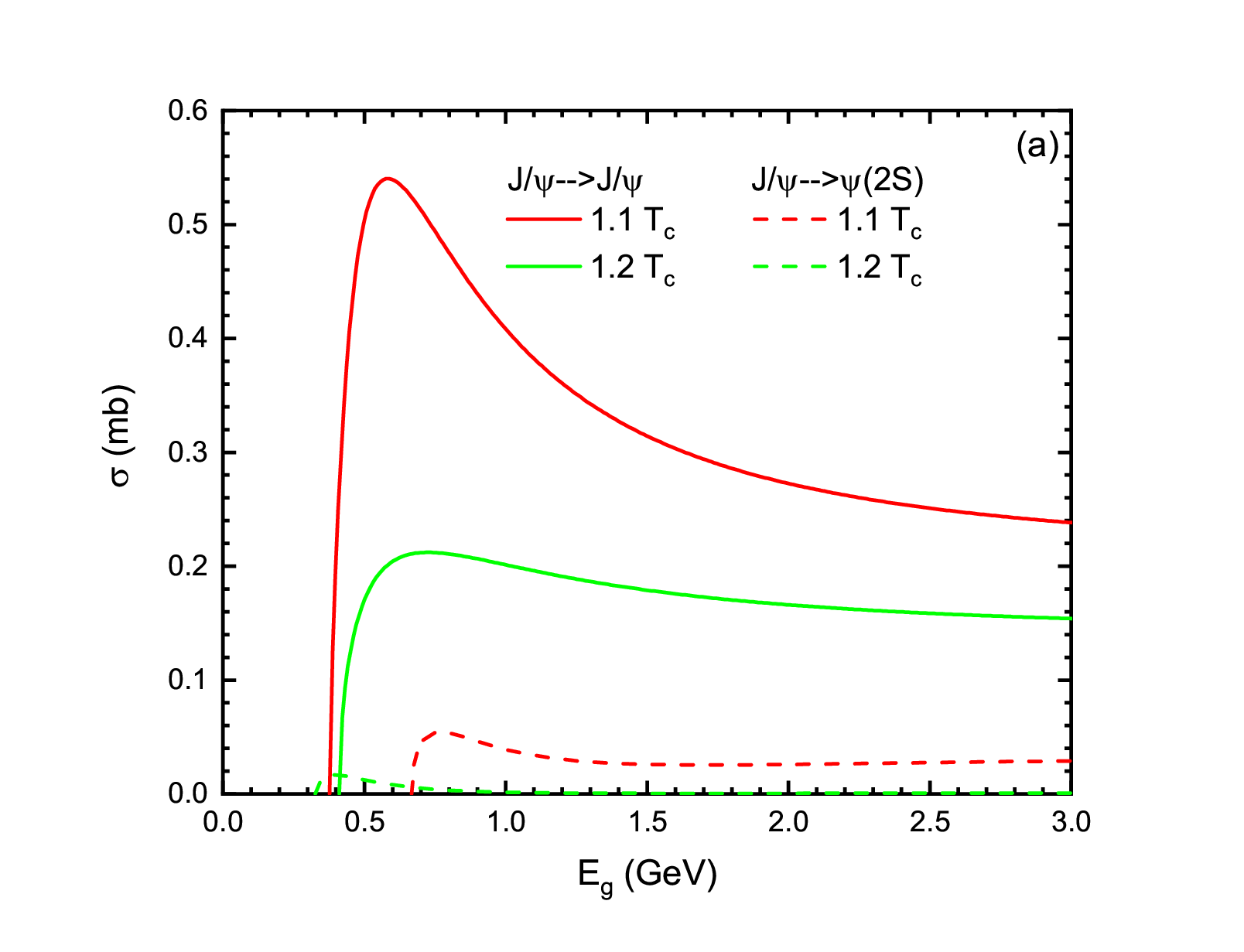}
	\vspace{-0.05cm}
    \includegraphics[width=1.0\columnwidth]{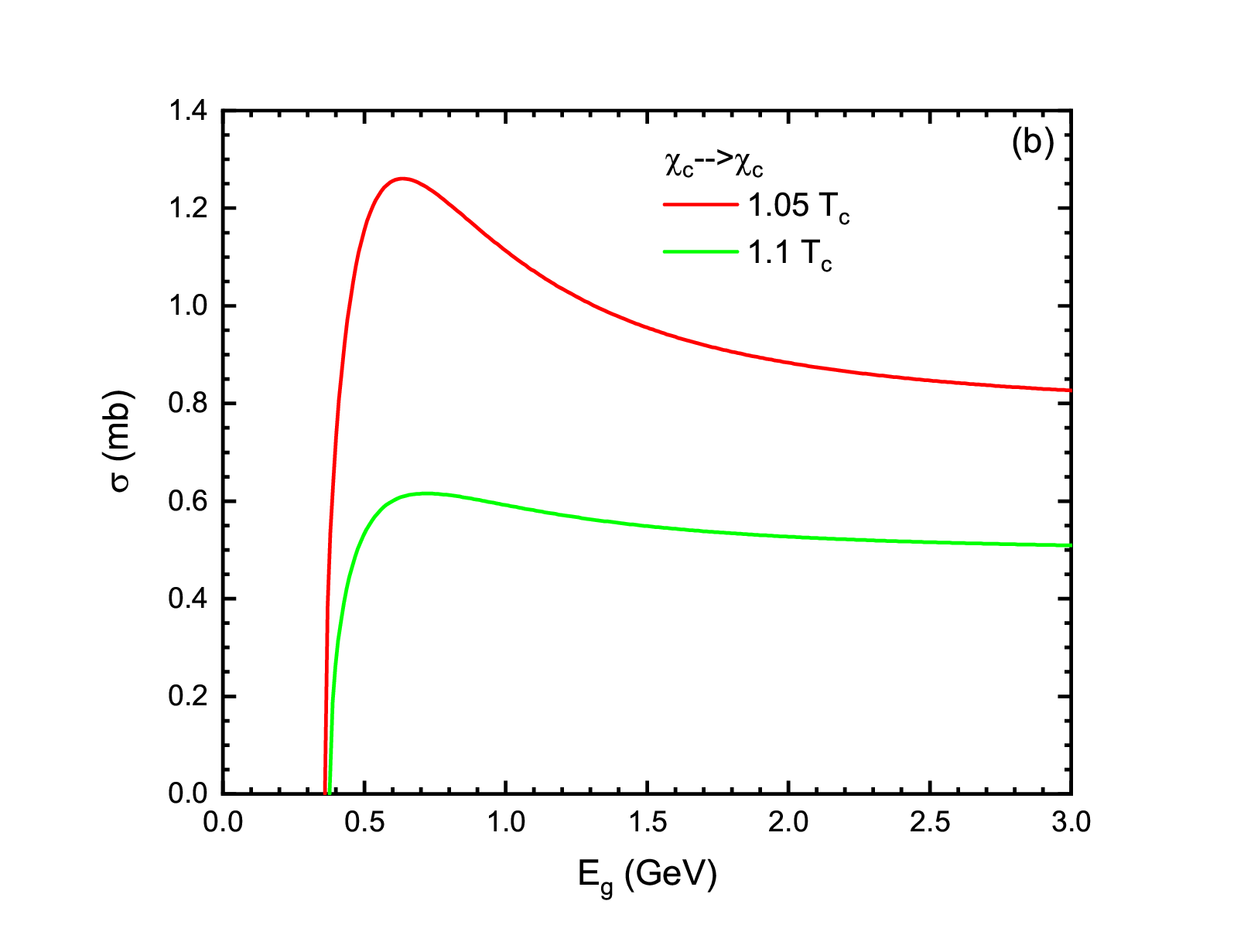}
	\vspace{-0.05cm}
	\caption{Cross sections for the second-order transitions between different charmonium bound states at finite temperatures. (a) $g+J/\psi\to g+J/\psi)$ (elastic collisions) and $g+J/\psi\to g+\psi(2S)$ (excitation). (b) $g+\chi_c\to g+\chi_c$ (elastic collision).}
	\label{figs_charmonia_tran_cross_sections}
\end{figure}

\begin{figure}[!t]
    \includegraphics[width=1.0\columnwidth]{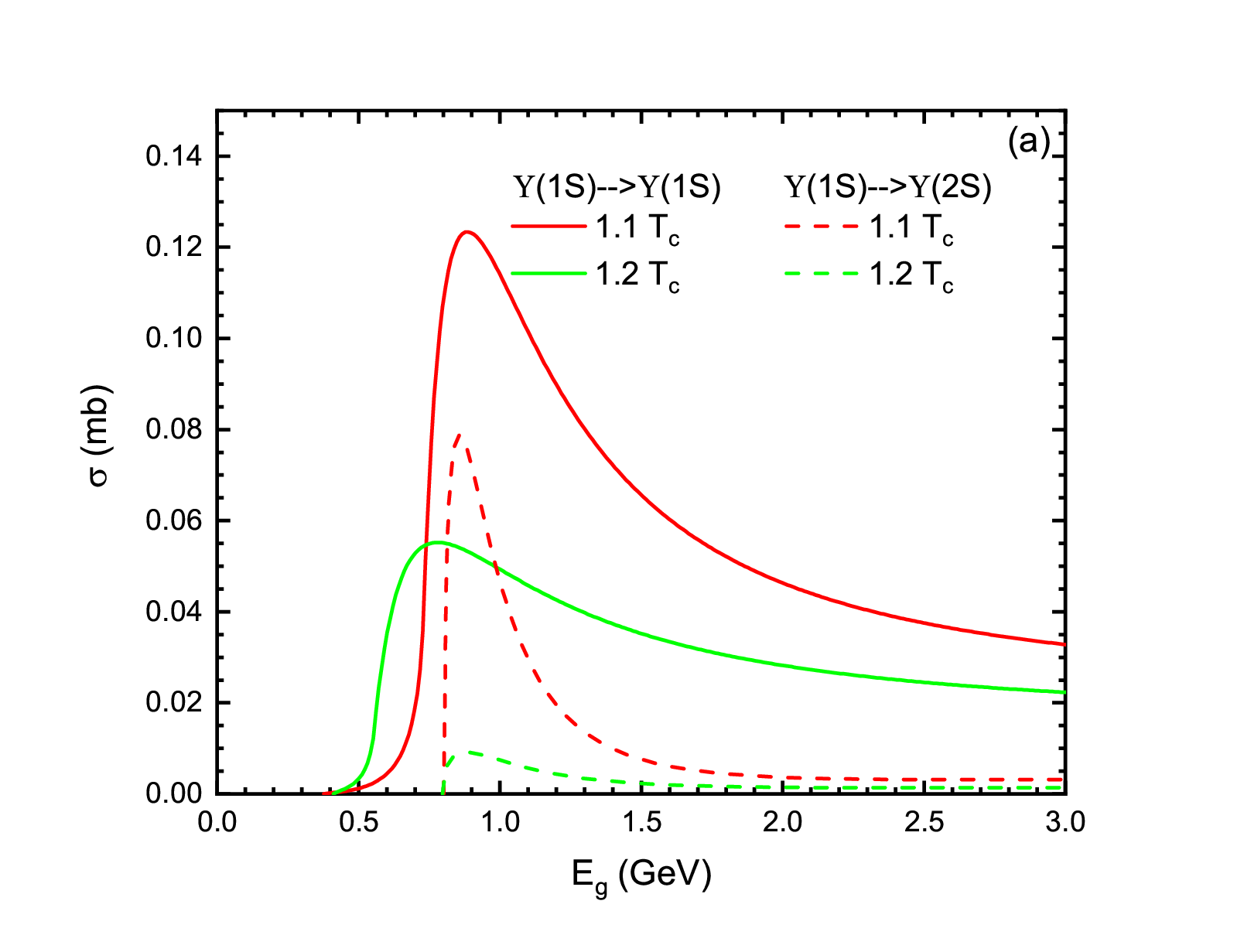}
	\vspace{-0.05cm}
    \includegraphics[width=1.0\columnwidth]{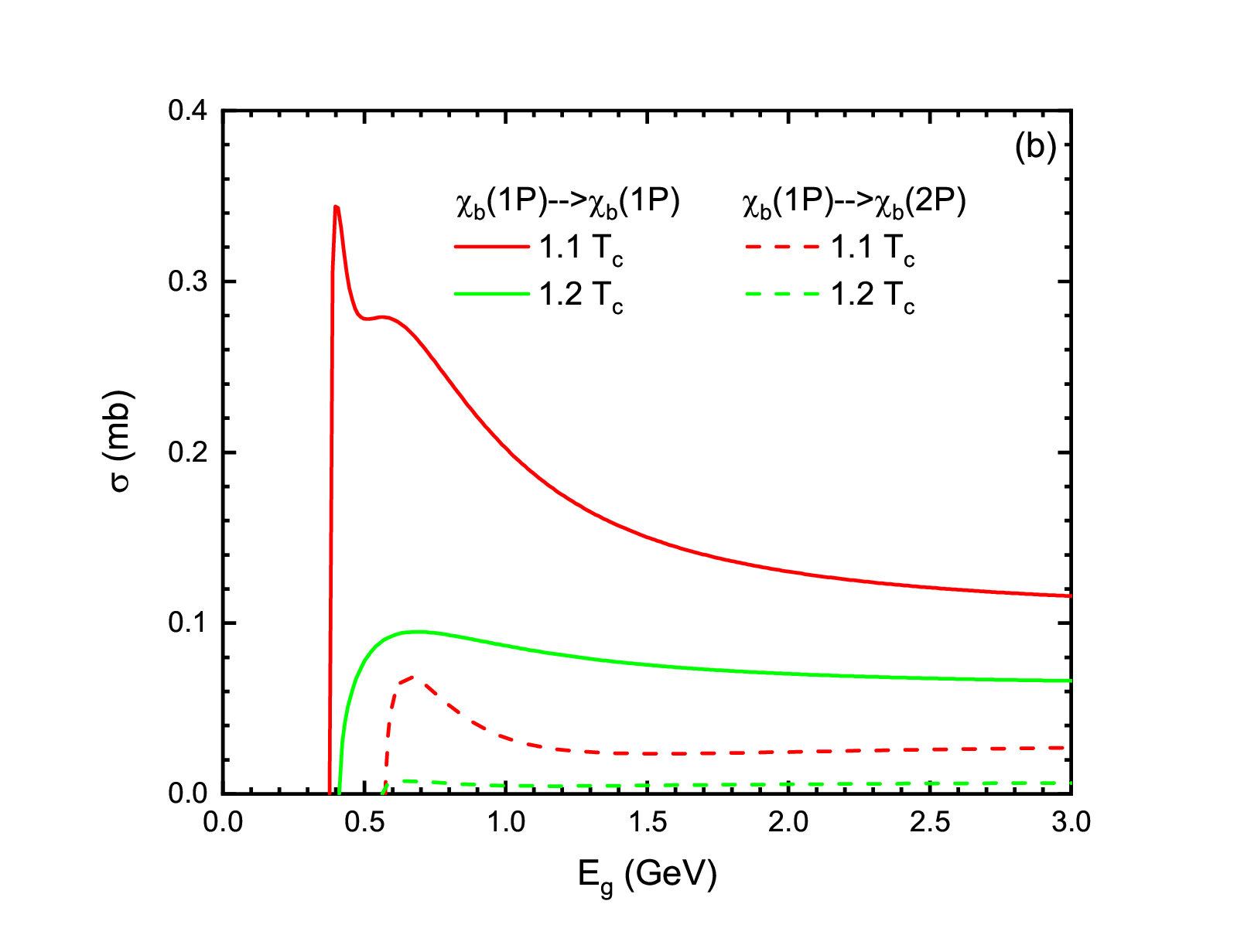}
	\vspace{-0.05cm}
	\caption{Cross sections for the second-order transitions between different bottomonium bound states at finite temperatures. (a) $g+\Upsilon(1S)\to g+\Upsilon(1S)$ (elastic collisions) and $g+\Upsilon(1S)\to g+\Upsilon(2S)$ (excitation). (b) $g+\chi_b(1P)\to g+\chi_b(1P)$ (elastic collision) and $g+\chi_b(1P)\to g+\chi_b(2P)$ (excitation).}
	\label{figs_bottomonia_tran_cross_sections}
\end{figure}

The computed second-order transition (elastic scattering or excitation) cross sections are presented in Fig.~\ref{figs_charmonia_tran_cross_sections} and Fig.~\ref{figs_bottomonia_tran_cross_sections} for charmonia and bottomonia, respectively. These cross sections share the same feature that they rise fast with incident gluon energy and peak near the threshold, which is followed by a decline, and gradually level off at large energies. In terms of magnitude, these elastic or excitation cross sections are much smaller than the total NLO dissociation cross sections, but are comparable to the component $\sigma^{(a+b)}$ (cf. Fig.~\ref{figs_charmonia_diss_cross_section}(a) and Fig.~\ref{figs_bottomonia_diss_cross_section}(a)) arising from diagrams (a) and (b) in Fig.~\ref{figs[a]+[b]}. The latter is not surprising since underlying all these diagrams is the same exchange of a color octet state. The cross sections for transition between $p$-wave bound states are generally larger than those involving $s$-wave bound states.

The second-order transition rates are computed using Eq.~(\ref{dissocation_rates}).  The numerical results for these rates are displayed in Fig.~\ref{figs_tran_rate_Charmonia} and Fig.~\ref{figs_tran_rate_Bottomonia} for charmonia and bottomonia, respectively. Contrary to the NLO dissociation rates growing with temperature (cf. Fig.~\ref{figs_diss_rate_Charmonia} and Fig.~\ref{figs_diss_rate_Bottomonia}), these rates decrease with temperature (except for the case of $\Upsilon(1S)$ elastic scattering which exhibits a moderate increase toward high temperatures). At rather low temperatures near $T_c$, these rates turn out to be comparable to the corresponding NLO dissociation rates, implying that it might be necessary to take these rates into account when interpreting the thermal spectral width of heavy quarkonium at low temperatures as computed in lattice QCD~\cite{Larsen:2019zqv}.

\begin{figure}[!t]
    \includegraphics[width=1.0\columnwidth]{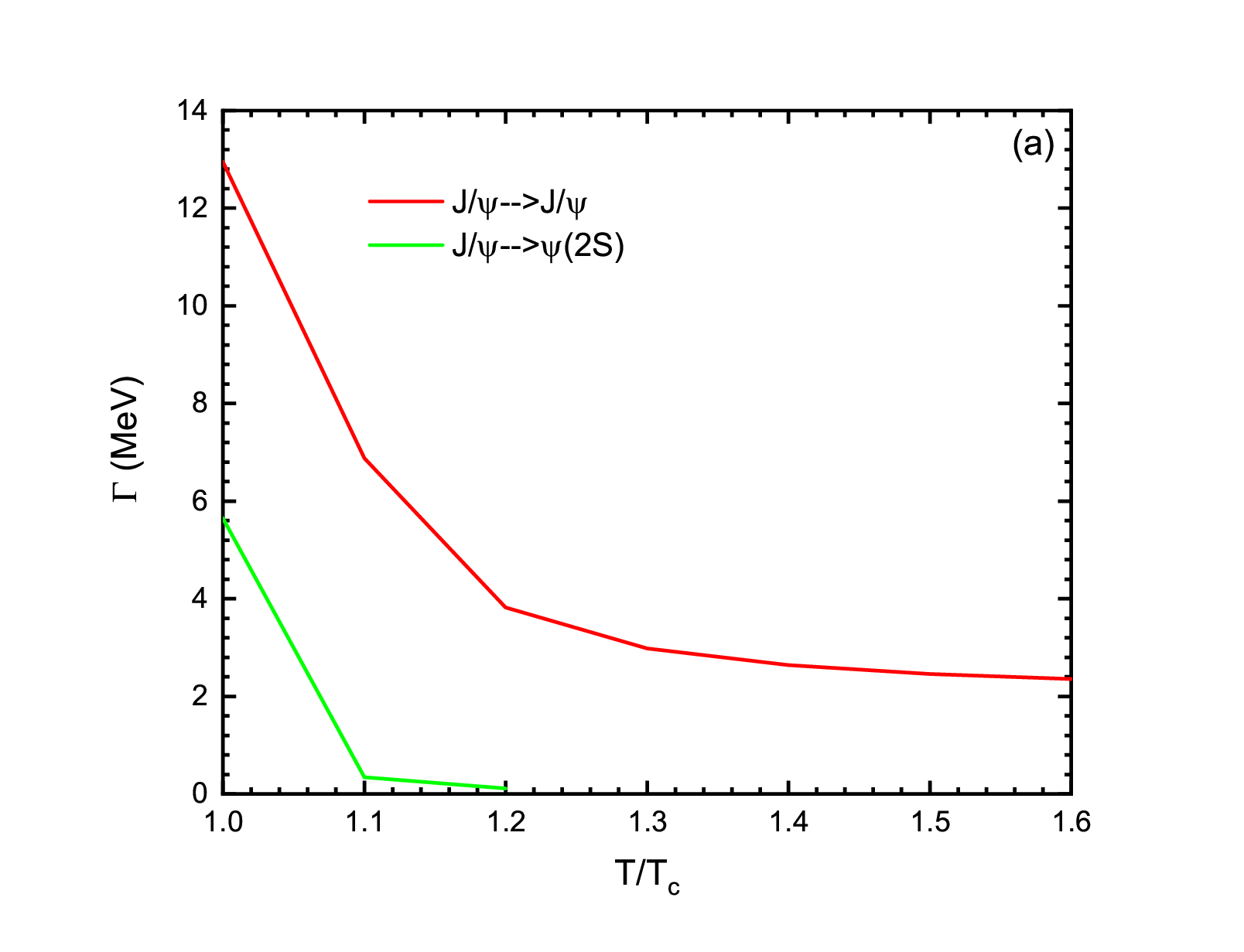}
	\vspace{-0.05cm}
    \includegraphics[width=1.0\columnwidth]{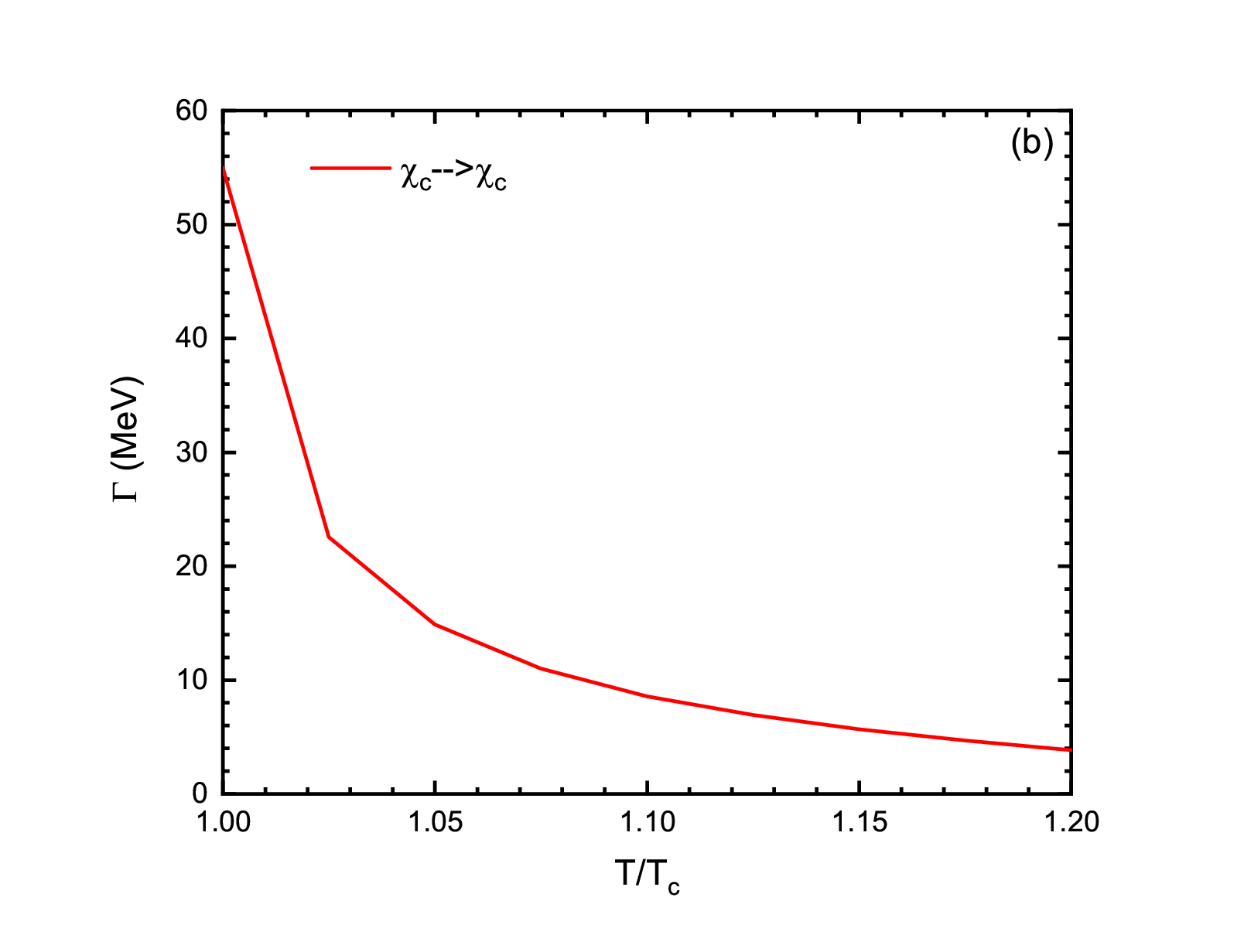}
	\vspace{-0.05cm}
	\caption{Second-order transition rates between different charmonium bound states at finite temperatures. (a) $g+J/\psi\to g+J/\psi)$ (elastic collisions) and $g+J/\psi\to g+\psi(2S)$ (excitation). (b) $g+\chi_c\to g+\chi_c$ (elastic collision). }
	\label{figs_tran_rate_Charmonia}
\end{figure}

\begin{figure}[!t]
    \includegraphics[width=1.0\columnwidth]{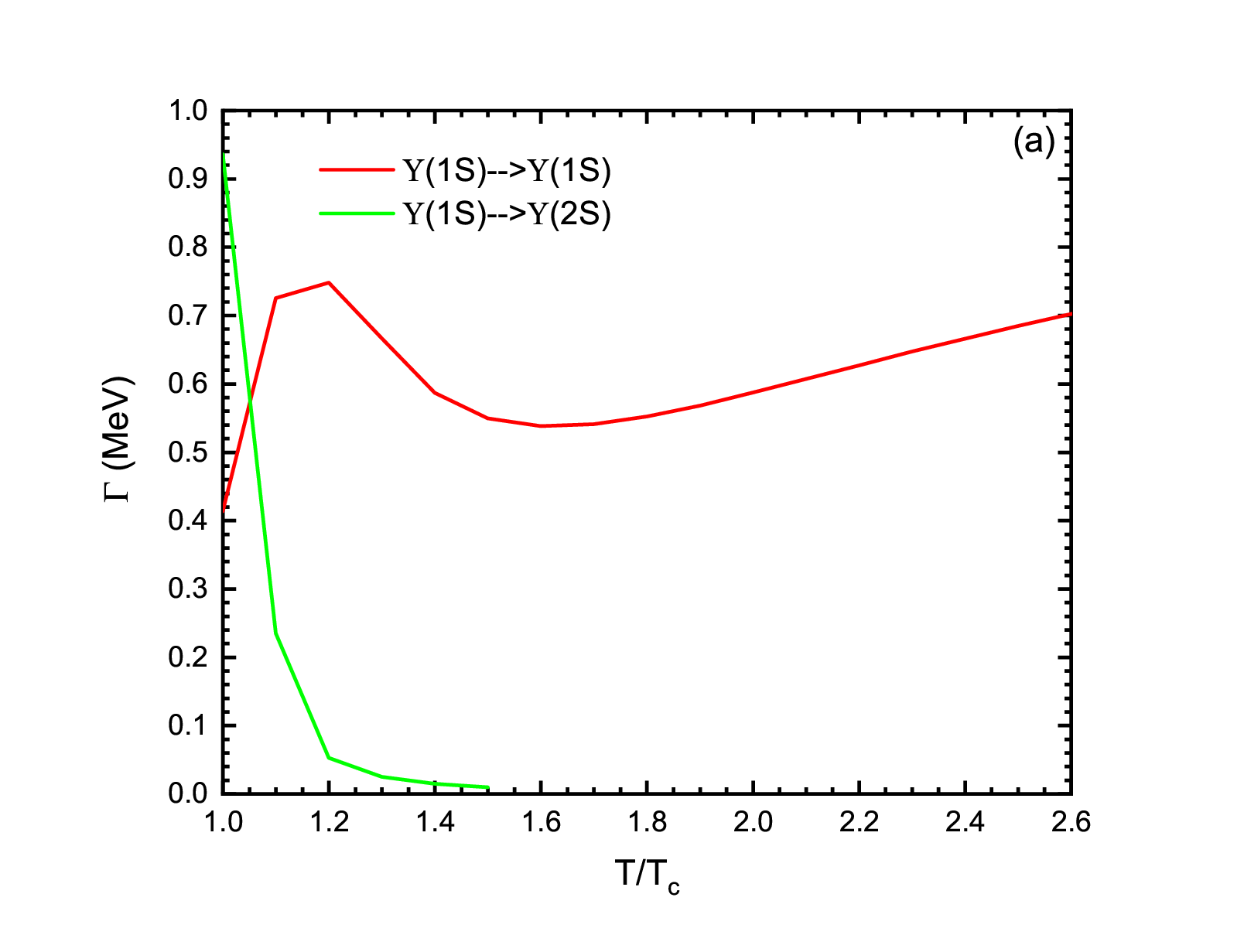}
	\vspace{-0.05cm}
    \includegraphics[width=1.0\columnwidth]{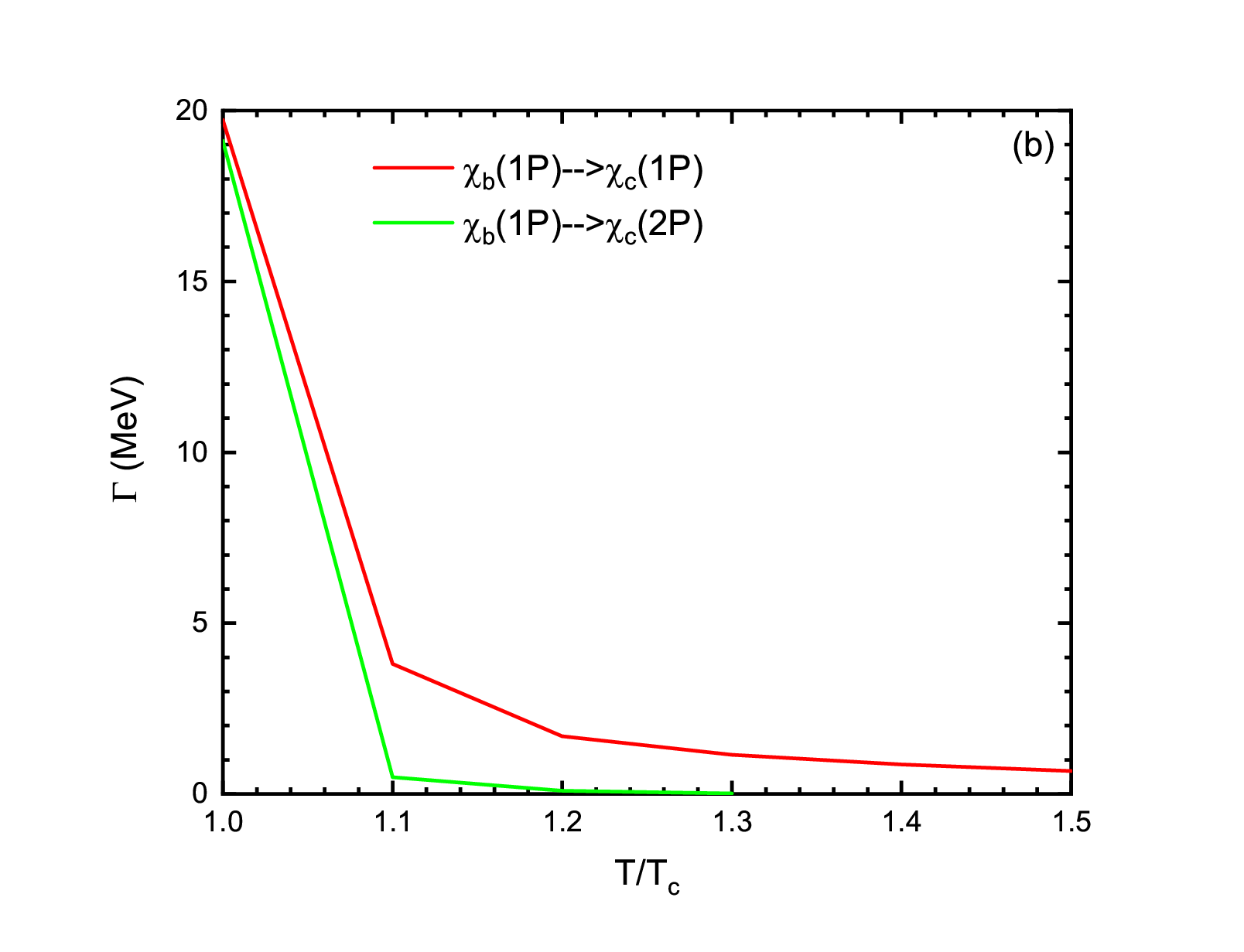}
	\vspace{-0.05cm}
	\caption{Second-order transition rates between different bottomonium bound states at finite temperatures. (a) $g+\Upsilon(1S)\to g+\Upsilon(1S)$ (elastic collisions) and $g+\Upsilon(1S)\to g+\Upsilon(2S)$ (excitation). (b) $g+\chi_b(1P)\to g+\chi_b(1P)$ (elastic collision) and $g+\chi_b(1P)\to g+\chi_b(2P)$ (excitation).}
	\label{figs_tran_rate_Bottomonia}
\end{figure}

%%%%%%%%%%%%%%%%%%%%%%%%%%%%%%%%%%%%%%%%%%%%%%%%%%%%%%%%%%%%%%%%
\section{Summary}
\label{sec_sum}
%%%%%%%%%%%%%%%%%%%%%%%%%%%%%%%%%%%%%%%%%%%%%%%%%%%%%%%%%%%%%%%%

In this work, we have presented a systematic study of the parton-induced second-order dissociation and transition processes for heavy quarkonia in the QGP within the approach of quantum mechanical perturbation theory, utilizing the effective chromo-electric dipole coupling of the heavy quarkonium with external gluons and a Hamiltonian formulation of QCD vertices. In this approach, the NLO dissociation and transition cross sections have been derived directly from the dynamical scattering point of view, which systematically incorporate the bound state wave functions and binding energies. This is in contrast to the indirect extraction of the cross sections from the imaginary heavy quark potential as done in~\cite{Brambilla:2013dpa} and also goes beyond the quasifree treatment~\cite{Grandchamp:2001pf,Rapp:2008tf}. Our approach also differs from the non-relativistic Bethe-Salpeter amplitude method~\cite{Hong:2018vgp} which considers only barely bound quarkonia with near-threshold energy, whereas our approach highlights the binding effects. Employing the bound state wave functions and binding energies calculated from an in-medium potential model, we have numerically evaluated the NLO dissociation and transition cross sections and rates for various charmonia and bottomonia at finite temperatures. Medium effects come into play through the temperature dependence of bound state wave functions (dipole size), binding energies and the thermal masses of light partons (which also cure the otherwise infrared and collinear divergences in evaluating cross sections). The NLO dissociation cross sections exhibit a rise and then a peculiar saturation as incident parton energy increases, as a result of the outgoing parton carrying away the excess energy, making them much more efficient in breaking up the bound states than the LO processes for energetic partons. Consequently, the NLO dissociation rates quickly take over from the LO counterparts at increasing temperatures. Finally we have also revealed that the second-order elastic scattering or excitation of bound state may induce significant widths for in-medium heavy quarkonium at temperatures not far from $T_c$.

The results presented here may be useful for phenomenological studies of heavy quarkonia transport in the relativistic heavy-ion collision systems. In particular, the parton-induced heavy quarkonium scattering amplitudes derived here might be of important use for more differential (beyond the simple rate equation approach~\cite{Rapp:2008tf}) and real-time simulations of heavy quarkonium dissociation and regeneration in the QGP in combination with the single heavy quark transport~\cite{Dong:2019byy,He:2021zej,He:2022ywp}.

\acknowledgments M. H. thanks Profs. Y.-G. Ma, G.-L. Ma and J.-H. Chen for the hospitality during his stay as a visiting scholar at Shanghai Research Center for Theoretical Nuclear Physics of Fudan University where the work was finalized. This work was supported by the National Natural Science Foundation of China (NSFC) under Grants No.12075122 (M. H.) and No.12147101 (via Shanghai Research Center for Theoretical Nuclear Physics).


\begin{thebibliography}{0}
\expandafter\ifx\csname natexlab\endcsname\relax\def\natexlab#1{#1}\fi
\expandafter\ifx\csname bibnamefont\endcsname\relax
  \def\bibnamefont#1{#1}\fi
\expandafter\ifx\csname bibfnamefont\endcsname\relax
  \def\bibfnamefont#1{#1}\fi
\expandafter\ifx\csname citenamefont\endcsname\relax
  \def\citenamefont#1{#1}\fi
\expandafter\ifx\csname url\endcsname\relax
  \def\url#1{\texttt{#1}}\fi
\expandafter\ifx\csname urlprefix\endcsname\relax\def\urlprefix{URL }\fi
\providecommand{\bibinfo}[2]{#2}
\providecommand{\eprint}[2][]{\url{#2}}

\end{thebibliography}


\begin{thebibliography}{99}

%\cite{Brambilla:2010cs}
\bibitem{Brambilla:2010cs}
N.~Brambilla, S.~Eidelman, B.~K.~Heltsley, R.~Vogt, G.~T.~Bodwin, E.~Eichten, A.~D.~Frawley, A.~B.~Meyer, R.~E.~Mitchell and V.~Papadimitriou, \textit{et al.}
%``Heavy Quarkonium: Progress, Puzzles, and Opportunities,''
Eur. Phys. J. C \textbf{71}, 1534 (2011).
%doi:10.1140/epjc/s10052-010-1534-9
%[arXiv:1010.5827 [hep-ph]].

%\cite{Rapp:2008tf}
\bibitem{Rapp:2008tf}
R.~Rapp, D.~Blaschke and P.~Crochet,
%``Charmonium and bottomonium production in heavy-ion collisions,''
Prog. Part. Nucl. Phys. \textbf{65}, 209-266 (2010).
%doi:10.1016/j.ppnp.2010.07.002
%[arXiv:0807.2470 [hep-ph]].


\bibitem{Braun-Munzinger:2009dzl}
P.~Braun-Munzinger and J.~Stachel,
%``Charmonium from Statistical Hadronization of Heavy Quarks \textendash{} a Probe for Deconfinement in the Quark-Gluon Plasma,''
Landolt-Bornstein \textbf{23}, 6-53 (2010).
%doi:10.1007/978-3-642-01539-7\_14
%[arXiv:0901.2500 [nucl-th]].


%\cite{Kluberg:2009wc}
\bibitem{Kluberg:2009wc}
L.~Kluberg and H.~Satz,
%``Color Deconfinement and Charmonium Production in Nuclear Collisions,''
Landolt-Bornstein \textbf{23}, 6-1 (2010).
%doi:10.1007/978-3-642-01539-7\_13.
%[arXiv:0901.3831 [hep-ph]].

%\cite{Mocsy:2013syh}
\bibitem{Mocsy:2013syh}
A.~Mocsy, P.~Petreczky and M.~Strickland,
%``Quarkonia in the Quark Gluon Plasma,''
Int. J. Mod. Phys. A \textbf{28}, 1340012 (2013).
%doi:10.1142/S0217751X13400125
%[arXiv:1302.2180 [hep-ph]].

%\cite{Andronic:2015wma}
\bibitem{Andronic:2015wma}
A.~Andronic, F.~Arleo, R.~Arnaldi, A.~Beraudo, E.~Bruna, D.~Caffarri, Z.~C.~del Valle, J.~G.~Contreras, T.~Dahms and A.~Dainese, \textit{et al.}
%``Heavy-flavour and quarkonium production in the LHC era: from proton\textendash{}proton to heavy-ion collisions,''
Eur. Phys. J. C \textbf{76}, no.3, 107 (2016).
%doi:10.1140/epjc/s10052-015-3819-5
%[arXiv:1506.03981 [nucl-ex]].


%\cite{Rothkopf:2019ipj}
\bibitem{Rothkopf:2019ipj}
A.~Rothkopf,
%``Heavy Quarkonium in Extreme Conditions,''
Phys. Rept. \textbf{858}, 1-117 (2020).
%doi:10.1016/j.physrep.2020.02.006
%[arXiv:1912.02253 [hep-ph]].


%\cite{Zhao:2020jqu}
\bibitem{Zhao:2020jqu}
J.~Zhao, K.~Zhou, S.~Chen and P.~Zhuang,
%``Heavy flavors under extreme conditions in high energy nuclear collisions,''
Prog. Part. Nucl. Phys. \textbf{114}, 103801 (2020).
%doi:10.1016/j.ppnp.2020.103801
%[arXiv:2005.08277 [nucl-th]].

%\cite{Andronic:2024oxz}
\bibitem{Andronic:2024oxz}
A.~Andronic, P.~B.~Gossiaux, P.~Petreczky, R.~Rapp, M.~Strickland, J.~P.~Blaizot, N.~Brambilla, P.~Braun-Munzinger, B.~Chen and S.~Delorme, \textit{et al.}
%``Comparative Study of Quarkonium Transport in Hot QCD Matter,''
arXiv:2402.04366 [nucl-th].


%\cite{Matsui:1986dk}
\bibitem{Matsui:1986dk}
T.~Matsui and H.~Satz,
%``$J/\psi$ Suppression by Quark-Gluon Plasma Formation,''
Phys. Lett. B \textbf{178}, 416-422 (1986).
%doi:10.1016/0370-2693(86)91404-8


%\cite{Digal:2001ue}
\bibitem{Digal:2001ue}
S.~Digal, P.~Petreczky and H.~Satz,
%``Quarkonium feed down and sequential suppression,''
Phys. Rev. D \textbf{64}, 094015 (2001).
%doi:10.1103/PhysRevD.64.094015
%[arXiv:hep-ph/0106017 [hep-ph]].

%\cite{Kharzeev:1994pz}
\bibitem{Kharzeev:1994pz}
D.~Kharzeev and H.~Satz,
%``Quarkonium interactions in hadronic matter,''
Phys. Lett. B \textbf{334}, 155-162 (1994).
%doi:10.1016/0370-2693(94)90604-1


%\cite{Grandchamp:2001pf}
\bibitem{Grandchamp:2001pf}
L.~Grandchamp and R.~Rapp,
%``Thermal versus direct J/Psi production in ultrarelativistic heavy ion collisions,''
Phys. Lett. B \textbf{523}, 60-66 (2001).
%doi:10.1016/S0370-2693(01)01311-9
%[arXiv:hep-ph/0103124 [hep-ph]].

%\cite{Braun-Munzinger:2000csl}
\bibitem{Braun-Munzinger:2000csl}
P.~Braun-Munzinger and J.~Stachel,
%``(Non)thermal aspects of charmonium production and a new look at J / psi suppression,''
Phys. Lett. B \textbf{490}, 196-202 (2000).
%doi:10.1016/S0370-2693(00)00991-6
%[arXiv:nucl-th/0007059 [nucl-th]].


%\cite{Thews:2000rj}
\bibitem{Thews:2000rj}
R.~L.~Thews, M.~Schroedter and J.~Rafelski,
%``Enhanced $J/\psi$ production in deconfined quark matter,''
Phys. Rev. C \textbf{63}, 054905 (2001).
%doi:10.1103/PhysRevC.63.054905
%[arXiv:hep-ph/0007323 [hep-ph]].


%\cite{Grandchamp:2003uw}
\bibitem{Grandchamp:2003uw}
L.~Grandchamp, R.~Rapp and G.~E.~Brown,
%``In medium effects on charmonium production in heavy ion collisions,''
Phys. Rev. Lett. \textbf{92}, 212301 (2004).
%doi:10.1103/PhysRevLett.92.212301
%[arXiv:hep-ph/0306077 [hep-ph]].


%\cite{Zhao:2010nk}
\bibitem{Zhao:2010nk}
X.~Zhao and R.~Rapp,
%``Charmonium in Medium: From Correlators to Experiment,''
Phys. Rev. C \textbf{82}, 064905 (2010).
%doi:10.1103/PhysRevC.82.064905
%[arXiv:1008.5328 [hep-ph]].

%\cite{Song:2011nu}
\bibitem{Song:2011nu}
T.~Song, K.~C.~Han and C.~M.~Ko,
%``Bottomonia suppression in heavy-ion collisions,''
Phys. Rev. C \textbf{85}, 014902 (2012).
%doi:10.1103/PhysRevC.85.014902
%[arXiv:1109.6691 [nucl-th]].

%\cite{Strickland:2011mw}
\bibitem{Strickland:2011mw}
M.~Strickland,
%``Thermal $\upsilon_{1s}$ and chi\_b1 suppression in $\sqrt{s_{NN}}=2.76$ TeV Pb-Pb collisions at the LHC,''
Phys. Rev. Lett. \textbf{107}, 132301 (2011).
%doi:10.1103/PhysRevLett.107.132301
%[arXiv:1106.2571 [hep-ph]].

%\cite{Zhou:2014kka}
\bibitem{Zhou:2014kka}
K.~Zhou, N.~Xu, Z.~Xu and P.~Zhuang,
%``Medium effects on charmonium production at ultrarelativistic energies available at the CERN Large Hadron Collider,''
Phys. Rev. C \textbf{89}, no.5, 054911 (2014).
%doi:10.1103/PhysRevC.89.054911
%[arXiv:1401.5845 [nucl-th]].

%\cite{Brambilla:2016wgg}
\bibitem{Brambilla:2016wgg}
N.~Brambilla, M.~A.~Escobedo, J.~Soto and A.~Vairo,
%``Quarkonium suppression in heavy-ion collisions: an open quantum system approach,''
Phys. Rev. D \textbf{96}, no.3, 034021 (2017).
%doi:10.1103/PhysRevD.96.034021
%[arXiv:1612.07248 [hep-ph]].

%\cite{Brambilla:2020qwo}
\bibitem{Brambilla:2020qwo}
N.~Brambilla, M.~\'A.~Escobedo, M.~Strickland, A.~Vairo, P.~Vander Griend and J.~H.~Weber,
%``Bottomonium suppression in an open quantum system using the quantum trajectories method,''
JHEP \textbf{05}, 136 (2021).
%doi:10.1007/JHEP05(2021)136
%[arXiv:2012.01240 [hep-ph]].


%\cite{Yao:2021lus}
\bibitem{Yao:2021lus}
X.~Yao,
%``Open quantum systems for quarkonia,''
Int. J. Mod. Phys. A \textbf{36}, no.20, 2130010 (2021).
%doi:10.1142/S0217751X21300106
%[arXiv:2102.01736 [hep-ph]].


%\cite{Wu:2024gil}
\bibitem{Wu:2024gil}
B.~Wu and R.~Rapp,
%``Charmonium Transport in Ultra-Relativistic Heavy-Ion Collisions at the LHC,''
[arXiv:2404.09881 [nucl-th]].


%\cite{Peskin:1979va}
\bibitem{Peskin:1979va}
M.~E.~Peskin,
%``Short Distance Analysis for Heavy Quark Systems. 1. Diagrammatics,''
Nucl. Phys. B \textbf{156}, 365-390 (1979).
%doi:10.1016/0550-3213(79)90199-8


%\cite{Bhanot:1979vb}
\bibitem{Bhanot:1979vb}
G.~Bhanot and M.~E.~Peskin,
%``Short Distance Analysis for Heavy Quark Systems. 2. Applications,''
Nucl. Phys. B \textbf{156}, 391-416 (1979).
%doi:10.1016/0550-3213(79)90200-1
%443 citations counted in INSPIRE as of 22 Apr 2024


%\cite{Oh:2001rm}
\bibitem{Oh:2001rm}
Y.~s.~Oh, S.~Kim and S.~H.~Lee,
%``Quarkonium hadron interactions in QCD,''
Phys. Rev. C \textbf{65}, 067901 (2002).
%doi:10.1103/PhysRevC.65.067901
%[arXiv:hep-ph/0111132 [hep-ph]].


%\cite{Chen:2017jje}
\bibitem{Chen:2017jje}
S.~Chen and M.~He,
%``Gluo-dissociation of heavy quarkonium in the quark-gluon plasma reexamined,''
Phys. Rev. C \textbf{96}, no.3, 034901 (2017).
%doi:10.1103/PhysRevC.96.034901
%[arXiv:1705.10110 [nucl-th]].


%\cite{Wong:2004zr}
\bibitem{Wong:2004zr}
C.~Y.~Wong,
%``Heavy quarkonia in quark-gluon plasma,''
Phys. Rev. C \textbf{72}, 034906 (2005).
%doi:10.1103/PhysRevC.72.034906
%[arXiv:hep-ph/0408020 [hep-ph]].


%\cite{Brezinski:2011ju}
\bibitem{Brezinski:2011ju}
F.~Brezinski and G.~Wolschin,
%``Gluodissociation and Screening of $\upsilon$ States in PbPb Collisions at $\sqrt{s_{NN}}=2.76$ TeV,''
Phys. Lett. B \textbf{707}, 534-538 (2012).
%doi:10.1016/j.physletb.2012.01.012
%[arXiv:1109.0211 [hep-ph]].

%\cite{Liu:2013kkg}
\bibitem{Liu:2013kkg}
Y.~Liu, C.~M.~Ko and T.~Song,
%``Gluon dissociation of J/\ensuremath{\psi} beyond the dipole approximation,''
Phys. Rev. C \textbf{88}, no.6, 064902 (2013).
%doi:10.1103/PhysRevC.88.064902
%[arXiv:1307.4427 [hep-ph]].


%\cite{Riek:2010py}
\bibitem{Riek:2010py}
F.~Riek and R.~Rapp,
%``Selfconsistent Evaluation of Charm and Charmonium in the Quark-Gluon Plasma,''
New J. Phys. \textbf{13}, 045007 (2011).
%doi:10.1088/1367-2630/13/4/045007
%[arXiv:1012.0019 [nucl-th]].


%\cite{Du:2017qkv}
\bibitem{Du:2017qkv}
X.~Du, M.~He and R.~Rapp,
%``Color Screening and Regeneration of Bottomonia in High-Energy Heavy-Ion Collisions,''
Phys. Rev. C \textbf{96}, no.5, 054901 (2017).
%doi:10.1103/PhysRevC.96.054901
%[arXiv:1706.08670 [hep-ph]].


%\cite{Song:2005yd}
\bibitem{Song:2005yd}
T.~Song and S.~H.~Lee,
%``Quarkonium-hadron interactions in perturbative QCD,''
Phys. Rev. D \textbf{72}, 034002 (2005).
%doi:10.1103/PhysRevD.72.034002
%[arXiv:hep-ph/0501252 [hep-ph]].


%\cite{Hong:2018vgp}
\bibitem{Hong:2018vgp}
J.~Hong and S.~H.~Lee,
%``Quarkonium dissociation in perturbative QCD,''
Phys. Rev. C \textbf{99}, no.3, 034905 (2019).
%doi:10.1103/PhysRevC.99.034905
%[arXiv:1811.07607 [nucl-th]].


%\cite{Laine:2006ns}
\bibitem{Laine:2006ns}
M.~Laine, O.~Philipsen, P.~Romatschke and M.~Tassler,
%``Real-time static potential in hot QCD,''
JHEP \textbf{03}, 054 (2007).
%doi:10.1088/1126-6708/2007/03/054
%[arXiv:hep-ph/0611300 [hep-ph]].


%\cite{Beraudo:2007ky}
\bibitem{Beraudo:2007ky}
A.~Beraudo, J.~P.~Blaizot and C.~Ratti,
%``Real and imaginary-time Q anti-Q correlators in a thermal medium,''
Nucl. Phys. A \textbf{806}, 312-338 (2008).
%doi:10.1016/j.nuclphysa.2008.03.001
%[arXiv:0712.4394 [nucl-th]].


%\cite{Brambilla:2008cx}
\bibitem{Brambilla:2008cx}
N.~Brambilla, J.~Ghiglieri, A.~Vairo and P.~Petreczky,
%``Static quark-antiquark pairs at finite temperature,''
Phys. Rev. D \textbf{78}, 014017 (2008).
%doi:10.1103/PhysRevD.78.014017
%[arXiv:0804.0993 [hep-ph]].


%\cite{Brambilla:2013dpa}
\bibitem{Brambilla:2013dpa}
N.~Brambilla, M.~A.~Escobedo, J.~Ghiglieri and A.~Vairo,
%``Thermal width and quarkonium dissociation by inelastic parton scattering,''
JHEP \textbf{05}, 130 (2013).
%doi:10.1007/JHEP05(2013)130
%[arXiv:1303.6097 [hep-ph]].


%\cite{Burnier:2014ssa}
\bibitem{Burnier:2014ssa}
Y.~Burnier, O.~Kaczmarek and A.~Rothkopf,
%``Static quark-antiquark potential in the quark-gluon plasma from lattice QCD,''
Phys. Rev. Lett. \textbf{114}, no.8, 082001 (2015).
%doi:10.1103/PhysRevLett.114.082001
%[arXiv:1410.2546 [hep-lat]].

%\cite{Bala:2021fkm}
\bibitem{Bala:2021fkm}
D.~Bala \textit{et al.} [HotQCD],
%``Static quark-antiquark interactions at nonzero temperature from lattice QCD,''
Phys. Rev. D \textbf{105}, no.5, 054513 (2022).
%doi:10.1103/PhysRevD.105.054513
%[arXiv:2110.11659 [hep-lat]].


%\cite{Bazavov:2023dci}
\bibitem{Bazavov:2023dci}
A.~Bazavov \textit{et al.} [HotQCD],
%``Unscreened forces in the quark-gluon plasma?,''
Phys. Rev. D \textbf{109}, no.7, 074504 (2024).
%doi:10.1103/PhysRevD.109.074504
%[arXiv:2308.16587 [hep-lat]].


%\cite{Yan:1980uh}
\bibitem{Yan:1980uh}
T.~M.~Yan,
%``Hadronic Transitions Between Heavy Quark States in Quantum Chromodynamics,''
Phys. Rev. D \textbf{22}, 1652 (1980).
%doi:10.1103/PhysRevD.22.1652


%\cite{Brambilla:2004jw}
\bibitem{Brambilla:2004jw}
N.~Brambilla, A.~Pineda, J.~Soto and A.~Vairo,
%``Effective Field Theories for Heavy Quarkonium,''
Rev. Mod. Phys. \textbf{77}, 1423 (2005).
%doi:10.1103/RevModPhys.77.1423
%[arXiv:hep-ph/0410047 [hep-ph]].


%\cite{Voloshin:2007dx}
\bibitem{Voloshin:2007dx}
M.~B.~Voloshin,
%``Charmonium,''
Prog. Part. Nucl. Phys. \textbf{61}, 455-511 (2008).
%doi:10.1016/j.ppnp.2008.02.001
%[arXiv:0711.4556 [hep-ph]].


%\cite{Sumino:2014qpa}
\bibitem{Sumino:2014qpa}
Y.~Sumino,
%``Understanding Interquark Force and Quark Masses in Perturbative QCD,''
arXiv:1411.7853 [hep-ph].


%\cite{Chen:2018dqg}
\bibitem{Chen:2018dqg}
S.~Chen and M.~He,
%``Heavy quarkonium dissociation by thermal gluons at next-to-leading order in the Quark\textendash{}Gluon Plasma,''
Phys. Lett. B \textbf{786}, 260-267 (2018).
%doi:10.1016/j.physletb.2018.09.056
%[arXiv:1805.06346 [nucl-th]].
%5 citations counted in INSPIRE as of 23 Apr 2024


%\cite{Larsen:2019zqv}
\bibitem{Larsen:2019zqv}
R.~Larsen, S.~Meinel, S.~Mukherjee and P.~Petreczky,
%``Excited bottomonia in quark-gluon plasma from lattice QCD,''
Phys. Lett. B \textbf{800}, 135119 (2020).
%doi:10.1016/j.physletb.2019.135119
%[arXiv:1910.07374 [hep-lat]].


%\cite{Yao:2018sgn}
\bibitem{Yao:2018sgn}
X.~Yao and B.~M\"uller,
%``Quarkonium inside the quark-gluon plasma: Diffusion, dissociation, recombination, and energy loss,''
Phys. Rev. D \textbf{100}, no.1, 014008 (2019).
%doi:10.1103/PhysRevD.100.014008
%[arXiv:1811.09644 [hep-ph]].


\bibitem{Peskin:1995}
M. E. Peskin and D. V. Schroeder, An Introduction to Quantum Field Theory, Addison-Wesley Publishing Company, 1995.


\bibitem{Sakurai-AQM:2008}
J. J. Sakurai, Advanced Quantum Mechanics, World Scientific, 2008.


\bibitem{Schwartz-QFT:2014}
M. D. Schwartz, Quantum Field Theory and the Standard Model, Cambridge University Press, 2014.


\bibitem{Greiner:1998}
W. Greiner, Quantum Mechanics: Special Chapters, Springer, 1998.


\bibitem{Gottfried-Yan:2003}
K. Gottfried and T.-M. Yan, Quantum Mechanics: Fundamentals, Springer, 2003.


\bibitem{Karsch:1987pv}
  F.~Karsch, M.~T.~Mehr and H.~Satz,
  %``Color Screening and Deconfinement for Bound States of Heavy Quarks,''
  Z.\ Phys.\ C {\bf 37}, 617 (1988).


%\cite{Burnier:2015tda}
\bibitem{Burnier:2015tda}
Y.~Burnier, O.~Kaczmarek and A.~Rothkopf,
%``Quarkonium at finite temperature: Towards realistic phenomenology from first principles,''
JHEP \textbf{12}, 101 (2015).
%doi:10.1007/JHEP12(2015)101
%[arXiv:1509.07366 [hep-ph]].


%\cite{Cao:2018ews}
\bibitem{Cao:2018ews}
S.~Cao, G.~Coci, S.~K.~Das, W.~Ke, S.~Y.~F.~Liu, S.~Plumari, T.~Song, Y.~Xu, J.~Aichelin and S.~Bass, \textit{et al.}
%``Toward the determination of heavy-quark transport coefficients in quark-gluon plasma,''
Phys. Rev. C \textbf{99}, no.5, 054907 (2019).
%doi:10.1103/PhysRevC.99.054907
%[arXiv:1809.07894 [nucl-th]].


%\cite{Gossiaux:2008jv}
\bibitem{Gossiaux:2008jv}
P.~B.~Gossiaux and J.~Aichelin,
%``Towards an understanding of the RHIC single electron data,''
Phys. Rev. C \textbf{78}, 014904 (2008).
%doi:10.1103/PhysRevC.78.014904
%[arXiv:0802.2525 [hep-ph]].



%\cite{Zhao:2023ucp}
\bibitem{Zhao:2023ucp}
J.~Zhao, J.~Aichelin, P.~B.~Gossiaux and K.~Werner,
%``Heavy flavor as a probe of hot QCD matter produced in proton-proton collisions,''
Phys. Rev. D \textbf{109}, no.5, 054011 (2024).
%doi:10.1103/PhysRevD.109.054011
%[arXiv:2310.08684 [hep-ph]].


%\cite{Braaten:1991jj}
\bibitem{Braaten:1991jj}
E.~Braaten and M.~H.~Thoma,
%``Energy loss of a heavy fermion in a hot plasma,''
Phys. Rev. D \textbf{44}, 1298-1310 (1991).
%doi:10.1103/PhysRevD.44.1298


%\cite{Dong:2019byy}
\bibitem{Dong:2019byy}
X.~Dong, Y.~J.~Lee and R.~Rapp,
%``Open Heavy-Flavor Production in Heavy-Ion Collisions,''
Ann. Rev. Nucl. Part. Sci. \textbf{69}, 417-445 (2019).
%doi:10.1146/annurev-nucl-101918-023806
%[arXiv:1903.07709 [nucl-ex]].


%\cite{He:2021zej}
\bibitem{He:2021zej}
M.~He, B.~Wu and R.~Rapp,
%``Collectivity of J/\ensuremath{\psi} Mesons in Heavy-Ion Collisions,''
Phys. Rev. Lett. \textbf{128}, no.16, 162301 (2022).
%doi:10.1103/PhysRevLett.128.162301
%[arXiv:2111.13528 [nucl-th]].


%\cite{He:2022ywp}
\bibitem{He:2022ywp}
M.~He, H.~van Hees and R.~Rapp,
%``Heavy-quark diffusion in the quark\textendash{}gluon plasma,''
Prog. Part. Nucl. Phys. \textbf{130}, 104020 (2023).
%doi:10.1016/j.ppnp.2023.104020
%[arXiv:2204.09299 [hep-ph]].

\end{thebibliography}
\end{document}